\documentclass[prd,preprint,nofootinbib,superscriptaddress,showpacs]{revtex4-1}

\usepackage{amsmath}    
\usepackage{amssymb}
\usepackage{graphicx}   
\usepackage{verbatim}   
\usepackage{color}      
\usepackage{subfigure}  
\usepackage{hyperref}   
\usepackage{dcolumn}	
\usepackage{bm}		
\usepackage{multirow}
\usepackage{CJK}
\usepackage{cancel}

\newcommand{\figref}[1]{Fig.~\ref{#1}}

\newcommand{\tr}{\mathrm {tr}}
\begin{document}


\preprint{UK/13-01 \\
INT-PUB-13-002 \\
RIKEN-QHP-77
}

\begin{CJK*}{UTF8}{} 
\CJKfamily{gbsn}
\title{Strangeness and charmness content of the nucleon from overlap fermions on $2+1$-flavor domain-wall fermion configurations}

\collaboration{$\chi$QCD Collaboration}
\author{M.~Gong (宫明)}
\affiliation{Dept.\ of Physics and Astronomy, University of Kentucky, Lexington, KY 40506, USA}
\author{A.~Alexandru}
\affiliation{Dept.\ of Physics, George Washington University, Washington, DC 20052, USA}
\author{Y.~Chen (陈莹)}
\affiliation{Institute of High Energy Physics and Theoretical Physics Center for
Science Facilities, Chinese Academy of Sciences, Beijing 100049, China}
\author{T.~Doi}
\affiliation{Theoretical Research Division, Nishina Center, RIKEN, Wako 351-0198, Japan}
\author{S.J.~Dong}
\affiliation{Dept.\ of Physics and Astronomy, University of Kentucky, Lexington, KY 40506, USA}
\author{T.~Draper}
\affiliation{Dept.\ of Physics and Astronomy, University of Kentucky, Lexington, KY 40506, USA}
\author{W.~Freeman}
\affiliation{Dept.\ of Physics, George Washington University, Washington, DC 20052, USA}
\author{M.~Glatzmaier}
\affiliation{Dept.\ of Physics and Astronomy, University of Kentucky, Lexington, KY 40506, USA}
\author{A.~Li (李安意)}
\affiliation{Institute for Nuclear Theory, University of Washington, Seattle, WA 98195, USA}
\author{K.F.~Liu (刘克非)}
\affiliation{Dept.\ of Physics and Astronomy, University of Kentucky, Lexington, KY 40506, USA}
\author{Z.~Liu (刘朝峰)}
\affiliation{Institute of High Energy Physics and Theoretical Physics Center for
Science Facilities, Chinese Academy of Sciences, Beijing 100049, China}

\begin{abstract}
We present a calculation of the strangeness and charmness contents $\langle N |
\bar{s}s | N \rangle$ and $\langle N | \bar{c}c | N \rangle$ of the nucleon
from dynamical lattice QCD with $2+1$ flavors.  The calculation is performed
with overlap valence quarks on 2+1-flavor domain-wall fermion gauge
configurations. The configurations are generated by the RBC collaboration on a
$24^3\times 64$ lattice with sea-quark mass $am_l=0.005$, $am_s=0.04$, and
inverse lattice spacing $a^{-1}=1.73\,{\rm GeV}$.  Both actions have chiral
symmetry which is essential in avoiding contamination due to the operator
mixing with other flavors.  The nucleon propagator and the quark loops are both
computed with stochastic grid sources, while low-mode substitution and low-mode
averaging methods are used respectively which substantially improve the signal-to-noise ratio.
We obtain the strangeness matrix element $f_{T_{s}} = m_s
\langle N| \bar{s}s|N\rangle/M_N = 0.0334(62)$, and the charmness content
$f_{T_{c}} =m_c\langle N|\bar{c}c|N\rangle/M_N = 0.094(31)$ which is resolved
from zero by 3 $\sigma$ precision for the first time.
\end{abstract}

\maketitle

\end{CJK*}

\section{Introduction}  \label{intro}

The strangeness and charmness content of the nucleon are of fundamental
importance to our understanding of the sea-quark contribution to nucleon
structure.  In particular, the sea-quark contribution to the scalar $u/d$ quark
content $\langle N| \bar{u}u|N\rangle$ is crucial to furthering our
understanding of the pion-nucleon sigma term. In addition to their relevance to
nucleon structure, the strangeness and charmness content of the nucleon have
drawn recent interest due to their relevance in dark matter
searches~\cite{Falk:1998xj,Ellis:2008hf,Giedt:2009mr}.  One popular candidate
for dark matter is a weakly interacting massive particle (WIMP).  The scalar
(spin-independent) effective four-fermion interaction between the WIMP and the
quarks, such as the neutralino-nucleon scattering in the context of the minimal supersymmetric standard model,
is $\alpha_{3q_i} \bar{\chi}\chi \bar{q_i}q_i$~\cite{Falk:1998xj,Ellis:2008hf}.
Additionally, the scalar neutralino-nucleus coupling is found to be much larger
than the axial-vector coupling~\cite{jungman_supersymmetric_1996}.  In this
case, the scalar contribution to the total $\chi$-nucleus cross section is

\begin{equation}
\sigma = \frac{4 m_r^2}{\pi} {\left[ Z f_p + (A-Z) f_n \right]}^2,
\end{equation}
where $m_r$ is the reduced $\chi$-nucleus mass and 
\begin{equation}    \label{f_N}
\frac{f_N}{m_N} = \sum_{q=u,d,s} f_{T_q}^{N}\frac{\alpha_{3q}}{m_q} + 
         \sum_{Q=c,b,t} f_{T_Q}^{N}\frac{\alpha_{3Q}}{m_Q},
\end{equation}
where $N = p, n$, and
\begin{eqnarray}  \label{f_T}
f_{T_q} &=& \frac{m_q \left< N | \bar{q} q | N \right>}{m_N} \\
f_{T_Q} &=& \frac{m_Q \left< N | \bar{Q} Q | N \right>}{m_N} 
\end{eqnarray}
are the contributions for the light ($q$)and heavy ($Q$) quarks respectively.
From the trace anomaly for the nucleon mass
\begin{equation}   \label{n_mass}
m_N = \sum_{q=u,d,s} m_q \left< N | \bar{q} q | N \right> +  \sum_{Q=c,b,t} m_Q \left< N | \bar{Q} Q | N \right>
        - \frac{7\alpha_s}{8\pi} \left< N | G_{\mu\nu}G_{\mu\nu} | N \right>
\end{equation}
and the heavy-quark expansion~\cite{Shifman:1978zn}, one can relate the glue
condensate in the nucleon to the heavy-quark condensate.  Equation~(\ref{n_mass})
becomes
\begin{equation}
m_N = \sum_{q=u,d,s} m_q \left< N | \bar{q} q | N \right> + \frac{27}{2} m_Q \left< N | \bar{Q} Q | N \right>
\end{equation}
and Eq.~(\ref{f_N}) is written as
\begin{equation}  \label{f_N_a}
\frac{f_N}{m_N} = \sum_{q=u,d,s} f_{T_q}^{N}\frac{\alpha_{3q}}{m_q} + \frac{2}{27} \bar{f}_{T_Q}^{N} 
\sum_{Q=c,b,t} \frac{\alpha_{3Q}}{m_Q},
\end{equation}
where $\bar{f}_{T_Q}^{N} = 1- \sum_{q=u,d,s} f_{T_q}^{N}$. This expression is
most often used in the analysis for dark matter
searches~\cite{Falk:1998xj,Ellis:2008hf,Giedt:2009mr}. Since the couplings
$\alpha_{3q}$ and $\alpha_{3Q}$ contain many terms that are proportional to the
quark mass, e.g.\ through Higgs exchange,
we see from Eqs.~(\ref{f_N}) and (\ref{f_N_a}) that 
the total spin-independent neutralino-nucleus cross section is
mainly proportional to $f_{T_q}$ and $f_{T_Q}$. Thus, it is important to determine them
precisely.

At energy scales comparable to $\Lambda_{\mathrm{QCD}}$, perturbative
calculations of $f_{T_{s}}$ and $f_{T_{c}}$ in the nucleon are prohibitively
difficult due to the nonperturbative nature of QCD.
However, these sea-quark matrix elements are accessible to lattice
calculations.  The first lattice calculations of $f_{T_{s}}$ were done with
Wilson fermions on quenched lattices and with heavy dynamical Wilson fermion
configurations~\cite{Fukugita:1994ba,Dong:1995ec,Gusken:1998wy}.  Those
calculations gave relatively large values for $f_{T_{s}}$ on the order of $\sim
0.19(1)$~\cite{Dong:1995ec}.  More recent calculations with Wilson-clover
dynamical fermions also yield large values ($f_{T_{s}} \sim 0.1$--$0.46)$~\cite{Durr:2011mp,Horsley:2011wr,Babich:2010at}, whereas calculations
with fermions incorporating chiral symmetry result in much smaller $f_{T_{s}}$
on the order of a few percent~\cite{Takeda:2010cw,Toussaint:2009pz,
Engelhardt:2010zr,Engelhardt:2012gd,Freeman:2012ry,Dinter:2012tt,
Oksuzian:2012rzb,Junnarkar:2013ac}.

The large value for $f_{T_{s}}$ found using Wilson-type fermions is due to the
additive renormalization of the quark mass due to the lattice-spacing-dependent
chiral symmetry breaking.  As a consequence, there is mixing between the
$\bar{u}u$ and $\bar{d}d$ operators and the $\bar{s}s$
operator~\cite{Michael:2001bv} \cite{Takeda:2010cw}.  This leads to a subtraction term
from $\left< N | \bar{s} s | N \right>$ which is proportional to
the matrix element $\left< N|\bar{u}u+\bar{d}d|N\right>$. Since the latter
involves the valence contribution, the subtraction turns out to be large.  For
example, it is found that
\begin{equation}
y= \frac{2\left< N|\bar{s}s |N\right>}{\left<N|\bar{u}u+\bar{d}d|N\right>}
\end{equation}
is changed from $y=0.53(12)$ to $y=-0.28(33)$ after the subtractions for the
$N_f=2$ lattice with the nonperturbatively improved clover fermions~\cite{Michael:2001bv}.  Similarly,
$y=0.336(3)$ becomes $y=0.059(37)(28)$ after subtraction for the $N_f=2$, $32^3
\times 64$ lattice with Wilson-clover fermions~\cite{Bali:2011ks}.  An
alternative way of evaluating the strangeness matrix element is to apply the
Feynman-Hellman theorem and take the derivative with respect to $m_{\pi}^2$ and
$m_K^2$ instead of the strange quark mass~\cite{Young:2009zb}.  This approach
avoids the additive mass part of the subtraction~\cite{Takeda:2010cw} and leads
to a small
$f_{T_s} = 0.033(16)(4)(2)$~\cite{Young:2009zb}.
 
To avoid the large systematic errors caused by explicit chiral symmetry
breaking with Wilson-type fermions, we instead adopt overlap fermions with exact chiral symmetry on the lattice for the valence and the quark loop,
for which the quark mass receives no additive renormalization.

In addition to having small $O(a^2)$ discretization
errors~\cite{Dong:2000mr,Draper:2005mh}, the overlap fermion that we use for
the valence quarks in the nucleon can also be used for the light and charm
quarks in the loop insertion with small $O(m^2a^2)$
error~\cite{Liu:2002qu,Li:2010pw}. This allows us to calculate both $f_{T_s}$
and $f_{T_c}$.

For a heavy quark of flavor $Q$, it is shown~\cite{Shifman:1978zn} that, to leading order in the heavy quark expansion, 
the matrix element $m_Q \langle N | \bar{Q} Q | N \rangle$ is related to the glue condensate in the nucleon,
\begin{equation}  \label{gc}
\sigma_Q \equiv m_Q \langle N | \bar{Q} Q | N \rangle\rightarrow - \frac{\alpha_s}{12 \pi}\langle  N | G G |N \rangle.
\end{equation}
If $c, b,$ and $t$ quarks are all considered to be heavy and the light quark contributions are ignored, the combination of the trace anomaly and Eq.~(\ref{gc}) predicts 
\begin{equation}  \label{heavy_quark}
\sigma_Q = \frac{2}{27} m_N = 70~{\rm MeV}.
\end{equation}

It would be interesting to check this prediction by a direct lattice
calculation, since the heavy quarks provide a significant contribution to the
WIMP-on-nucleon cross section.  In previous calculations, the charmness content
of the nucleon $\left< N|\bar{c}c |N\right> = 0.056(27)$ has been obtained with
the MILC HISQ configurations~\cite{Freeman:2012ry} with a 2 $\sigma$ signal. In
this work, we calculate the strangeness $\langle N | \bar{s}s | N \rangle$ and
charmness content $\langle N | \bar{c}c | N \rangle$ of the nucleon in lattice
QCD with overlap valence quarks on 2+1-flavor domain-wall fermion gauge
configurations.

The calculation is done on a $24^3\times 64$ lattice with sea-quark mass
$m_l=0.005$, $m_s=0.04$, and lattice spacing $a^{-1}=1.73\,{\rm GeV}$.  We
obtain the strangeness content $f_{T_{s}} = m_s \langle N|
\bar{s}s|N\rangle/M_N = 0.0334(62) $ which has more than 5 $\sigma$ precision.
Compared with previous lattice calculations either with disconnected insertion
or via the Feynman-Hellman theorem~\cite{Durr:2011mp,Horsley:2011wr,Takeda:2010cw,Toussaint:2009pz,Engelhardt:2012gd,Freeman:2012ry,Dinter:2012tt,Oksuzian:2012rzb,Junnarkar:2013ac,Bali:2011ks,Young:2009zb},
the present calculation has the smallest
error.  We also obtain the charmness content $f_{T_{c}} =m_c\langle
N|\bar{c}c|N\rangle/M_N = 0.094(31)$. This is the first time the charmness has
been calculated with more than 3 $\sigma$ precision.

Our paper is organized as follows.
The overlap formulation is briefly summarized in Sec.~\ref{Sec:overlap}.
The technical details for calculating the nucleon two-point functions and the
quark loops are described in Sec.~\ref{Sec:2pt} and Sec.~\ref{Sec:loop}
respectively.  They are combined to calculate the disconnected three-point
functions in Sec.~\ref{Sec:3pt}.  Finally, the numerical results with chiral
extrapolation are presented in Sec.~\ref{Sec:results} and the conclusions are
given in Sec.~\ref{Sec:conclusion}.

\section{Overlap Fermions}
\label{Sec:overlap}

We adopt the overlap fermion formulation for the valence quarks in the nucleon
correlation functions as well as for the quark loops.  The inversion of overlap
fermions using deflation of low eigenmodes and the construction of meson and
nucleon two-point functions with low-mode substitution have been detailed
previously~\cite{Li:2010pw}.  Deflation with low eigenmodes and hypercubic smearing
speed up the inversion by a factor of $\sim 50$ for the $24^3 \times 64$
lattice that we use for this calculation and low-mode substitution (LMS)
improves the errors of the meson and nucleon correlators by a factor of $\sim$
3 -- 4~\cite{Li:2010pw}.  The overlap quark propagators are calculated on gauge
configurations with $2+1$-flavors of dynamical domain-wall fermions (DWF).  As
we mentioned in the Introduction, we adopt both the overlap and DWF fermion
formalisms since they preserve chiral symmetry via the Ginsparg-Wilson
relation.  Due to the high degree of chiral symmetry, the calculation of the
quark content is free of the problems that plague the Wilson-type fermions as
outlined in the above discussion. As an additional advantage, the $O(m^2 a^2)$
discretization errors are small for the overlap fermion.  This allows us to
compute the charmness contribution to the nucleon in addition to the
strangeness contribution.

The overlap operator~\cite{Neuberger:1997fp} is defined as
\begin{equation}
D_{ov}  (\rho) =   1 + \gamma_5 \varepsilon (\gamma_5 D_{\rm w}(\rho)),
\end{equation}
where $\varepsilon$ is the matrix sign function and $D_{\rm w}(\rho)$ is the
usual Wilson fermion operator, except with a negative mass parameter $- \rho =
1/2\kappa -4$ in which $\kappa_c < \kappa < 0.25$.  We set $\kappa = 0.2$ in
our calculation, corresponding to $\rho = 1.5$. The massive overlap Dirac
operator is defined as
\begin{eqnarray}
D_m &=& \rho D_{ov} (\rho) + m\, (1 - \frac{D_{ov} (\rho)}{2}) \nonumber\\
       &=& \rho + \frac{m}{2} + (\rho - \frac{m}{2})\, \gamma_5\, \varepsilon (\gamma_5 D_w(\rho)).
\end{eqnarray}
To accommodate the chiral transformation, it is usually convenient to use the
chirally regulated field $\hat{\psi} = (1 - \frac{1}{2} D_{ov}) \psi$ in lieu
of $\psi$ in the interpolation field and the currents.  This leads to an
effective propagator
\begin{equation}
G \equiv D_{\mathrm{eff}}^{-1} \equiv (1 - \frac{D_{ov}}{2}) D^{-1}_m = \frac{1}{D_c + m},
\end{equation}
where $D_c = \frac{\rho D_{ov}}{1 - D_{ov}/2}$ is chiral, i.e. $\{\gamma_5,
D_c\}=0$~\cite{Chiu:1998gp}.  It is worthwhile to point out that this effective
propagator has the same form as that in the continuum~\cite{Liu:2002qu}. In
other words, the inverse of the propagator is a chirally invariant massless
Dirac operator plus the quark mass term. As long as the $O(m^2 a^2)$ error is
small, this formulation is suitable for both light and heavy quarks.

We adopt the Zolotarev approximation to evaluate the matrix sign function. This
entails two nested conjugate gradient loops to calculate the propagator of the
overlap fermion. For each conjugate gradient loop, we use deflation with low
eigenmodes to speed up the inversion. The details are given in
Ref.~\cite{Li:2010pw}.  Due to the normality of $D_{ov}$,
i.e. $D_{ov}^{\dagger} D_{ov} = D_{ov} D_{ov}^{\dagger}$ and the
Ginsparg-Wilson relation $\{\gamma_5, D_{ov}\} = D_{ov}\gamma_5 D_{ov}$, the
eigenvalues of $D_{ov}$ are on a unit circle with the center at unity. The real
and chiral modes are at 0 and 2.  Others on the circle are paired with
conjugate eigenvalues. In other words, if $v_i$ is an eigenvector
\begin{equation}
D_{ov}v_i = \lambda_i v_i,
\end{equation}
then its conjugate partner $\gamma_5 v_i$ is also an eigenvector with eigenvalue $\lambda^*$,
\begin{equation}   \label{conjugate}
D_{ov}\gamma_5v_i = \lambda_i^*\gamma_5 v_i.
\end{equation}
To compute the quark propagator, we first find a few hundred pairs of the
lowest eigenvectors of the massless overlap operator in addition to the zero
modes.  Once we have obtained these lowest eigenvectors, we can solve the
high-mode part of the quark propagator by projecting out the low modes from the
source
\begin{equation}
D_{eff} G^H \eta = \left(1 - \sum_{i} \left( v_iv_i^\dagger  + \gamma_5v_iv_i^\dagger \gamma_5 \right) (1 - \frac{1}{2}\delta_{\lambda_i,0}) \right) \eta,
\end{equation}
where $\eta$ is the source vector and the factor $1 -
\frac{1}{2}\delta_{\lambda_i,0}$ takes care of the zero modes which are either
left-handed or right-handed.

In contrast, the low-mode part of the effective quark propagator can be
constructed with eigenvectors directly,
\begin{equation} \label{lowmode_propagator}
G^L =  \sum_{i}
\left[ \frac{(1 - \frac{\lambda_i}{2}) v_iv_i^\dagger }{\rho \lambda_i + m (1- \frac{\lambda_i}{2})}
+ \frac{ (1 - \frac{\lambda_i^*}{2})\gamma_5 v_i v_i^\dagger \gamma_5}{\rho \lambda_i^* + m (1- \frac{\lambda_i^*}{2})}\right]
(1 - \frac{1}{2}\delta_{\lambda_i,0}).
\end{equation}
For a given source $\eta$, the total effective quark propagator is
\begin{equation}  \label{HL}
G\eta = G^H\eta + G^L\eta.
\end{equation}
However, we should point out that since $G^L$ is constructed from the
eigenmodes rather than by inverting a source vector, we can compute the
any-to-any propagator for any source and sink location with little additional
computation.  We shall use this fact to carry out the low-mode substitution to
replace $G^L\eta$ in Eq.~(\ref{HL}) with $G^L$ for a source with given grid
points which greatly improves the nucleon correlator. This will be explained in
more detail in the next section.

\section{The Nucleon Two-Point Correlation Function}
\label{Sec:2pt}

\begin{figure}
\begin{center}
\includegraphics[width=2.5in]{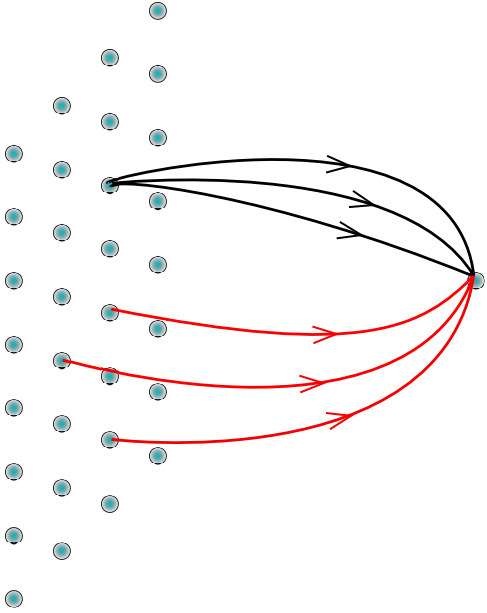}
\end{center}
\caption{
\label{diag.grid.nucleon}
(Color online) Diagram to illustrate the signal and noise of the nucleon
correlation function with a $Z_3$ noise grid source on a time slice. The upper
part with three quarks originating from the same spatial site is an example of
the gauge-invariant signal and the lower one is an example of the gauge-noninvariant
noise with three quarks originating from different spatial sites
which will be suppressed by gauge average and noise average.  }
\end{figure}

Various attempts have been made to improve the statistics of hadronic two-point
correlation functions such as using a smeared source, a volume source with
fixed gauge, and all-to-all propagators. The computation of all-to-all
propagators usually involves noise sources on different sites.  However, the
quark propagator from one site can be contaminated by those from neighboring
sites.  For example, when constructing the nucleon correlation function, the
three quark propagators may be from the same source site or from different
source sites.  The latter case is not gauge-invariant and will introduce noise
after averaging over a finite number of configurations.
\figref{diag.grid.nucleon} shows the gauge-invariant and the noninvariant
parts of the correlation function.

The signal-to-noise issue has been examined~\cite{Li:2010pw} for the connected
hadron correlators from the noise source on a time slice and it was found that
the noise wall source was worse than the point source for all mesons except the
pion. It is worse still for the nucleon. In this case, the signal-to-noise
ratio is
\begin{equation}
\frac{C(t, \vec{p} = 0)}{\sigma} \approx \sqrt{\frac{N}{V_3}} e^{-(m_N - 3/2 m_{\pi})t},
\end{equation}
where $N$ is the number of noises and $V_3$ is the three-volume of the time
slice. In addition to the usual exponential suppression in time, there is a
prefactor which reduces the signal-to-noise ratio further by a factor of
$\sqrt{V_3}$.  The situation can be ameliorated by reducing the noise
contamination from neighboring sites with less source points.  This introduces
the idea of a noise grid source with support on some uniformly spaced grid
points on a time slice, but this does not fundamentally alter the conclusion
that, given the same computer time, any noise source is worse than the point
source for the meson and nucleon.  These observations suggest a
new algorithm for the grid noise with low-mode substitution (LMS) to reduce the
variance from noise contamination while simultaneously addressing the low-mode correlation~\cite{Li:2010pw}.
The idea is to replace the low-mode part of the
quark propagator $G^L\eta$ estimated from noise sources by the exact $G^L$ in
Eq.~\eqref{lowmode_propagator} when all three quarks are in the low modes or
when two quarks are in the low modes and the third is in the high modes -- in
constructing the baryon correlators to reduce the noise contributions shown in
\figref{diag.grid.nucleon}.  It turns out that this LMS is quite successful. It
reverses the above-mentioned trend that a noise source is worse than a point
source and instead reduces the errors compared to the point
source~\cite{Li:2010pw}.

The $Z_3$ grid noise is
\begin{equation}   \label{noise_source}
\eta(x) = \sum_{i\in \mathcal{G}} \theta_i \delta_{x,i},
\end{equation}
where $\mathcal{G}$ is a sparse grid of lattice sites on a time slice, and
$\theta_i$ is the $Z_3$ random phase on site $i$ with the property
$\theta_i^3=1$ and $\langle \theta_i \theta_j \theta_k\rangle =
\delta_{ij}\delta_{jk}$.  Tests on the $32^3 \times 64$ lattice with a $Z_3$
noise on 64 evenly spaced grid points on a time slice and with LMS reveal that
for light quarks (pion masses at $200 - 300$ MeV) the errors of the meson and
nucleon masses can be reduced by a factor of 3 to 4 as compared to the point
source~\cite{Li:2010pw}.

Since the nucleon has a physical size of $\sim 0.6$--$0.8$ fm as deduced from
its axial and electromagnetic form factors, a smeared source for the nucleon
usually leads to a reduction of errors for the nucleon mass from the point
source. In view of this, we introduce a smeared-grid source to increase the
overlap with the nucleon ground-state wave function and diminish the
contribution from the radially excited states and the collateral $\pi N$
scattering states.  We adopt the gauge-invariant spatial Gaussian
smearing~\cite{Alexandrou:1992ti} on the grid source,
\begin{equation}
\eta^S(x^\prime) = S(x^\prime, x) \eta(x),
\end{equation}
where $S(x^\prime, x)$ is the smearing operator. By design, the smearing
operator $S(x^\prime, x)$ should produce a Gaussian distribution with a
Klein-Gordon propagator. It is computed as an iteration of many small smearing
steps,
\begin{equation}
S(x^\prime, x) = \left( 1 - \frac{3 w^2}{2 n} \right)^n \left[ 1 + \frac{w^2}{4 n - 6 w^2} \sum_{i=1}^3 \left( U_i(x^\prime,t) \delta_{x^\prime, x-\hat{i}} + U_i^\dagger(x^\prime-\hat{i},t) \delta_{x^\prime,x+\hat{i}} \right) \right]^{n},
\end{equation}
where $w$ is the input width for a Gaussian distribution and $n$ is the number of
smearing steps. The corresponding quark propagator with a smeared source is
\begin{eqnarray}  \label{smeared_propagator}
G(y,\eta^S) &=& D^{-1}(y,x^\prime) \eta^S(x^\prime) ,\nonumber\\
	&=& \sum_{i\in \mathcal{G}} \theta_i D^{-1}(y,x^\prime) S(x^\prime, x) \delta_{x,i}.
\end{eqnarray}

\begin{figure}[!htp]
\begin{center}
\includegraphics[width=5in]{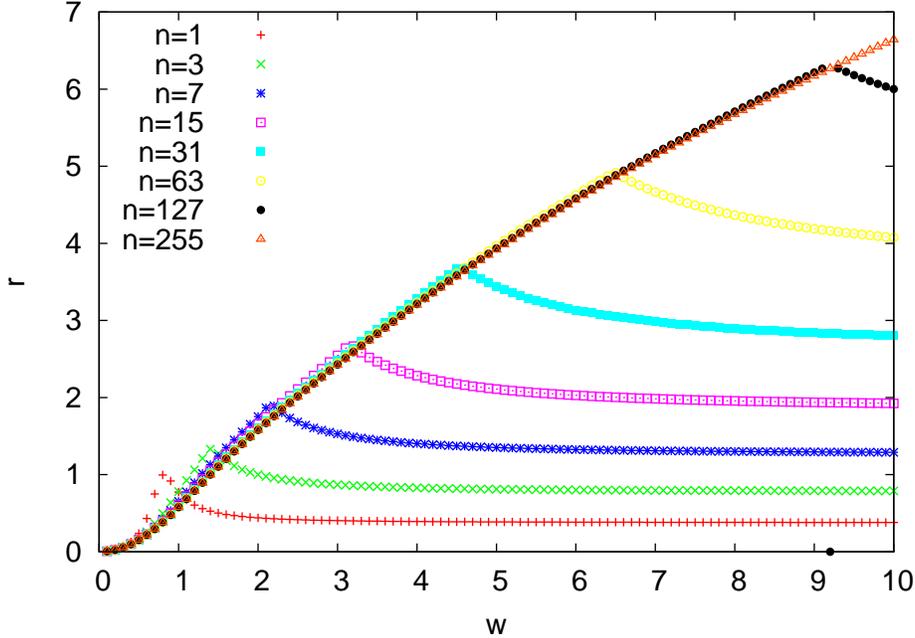}
\end{center}
\caption{
\label{plot.smear.size}
The plot shows the relation between the input width parameter $w$ and the
output radius of the gauge-invariant Gaussian smearing $r$ for different
iteration numbers $n$.  }
\end{figure}

To check if the actual width from the smearing procedure is consistent with the
input width parameter $w$, we first define the actual smearing size by
\begin{equation}
r = \sqrt{\frac{\sum_x x^2 \rho(x)}{\sum_x \rho(x)}},
\end{equation}
where, for each spatial position $x$, $\rho(x)$ is the norm over spin and color
of the smeared source vector which is created from a point source vector
$\delta(x)$. Then we plot $r$ vs $w$ for different step sizes $n$ in
\figref{plot.smear.size}.  We see that for a given $n$, there is a range of $w$
where the resultant smearing size $r$ has a nearly linear relation with
$w$. Beyond that range, $r$ flattens off as $w$ increases. As $n$ increases,
the range for the near-linear relation expands.  Even though we do not have
evidence that the smeared source, when its $r$ is much smaller than the input
$w$, leads to ill physical effects, we think it is safe to keep the linear
relationship. Consequently, we will use sufficient $n$ for a prescribed width
$w$ so that $r$ and $w$ are in the linear range.

\begin{figure}
\begin{center}
\includegraphics[width=5in]{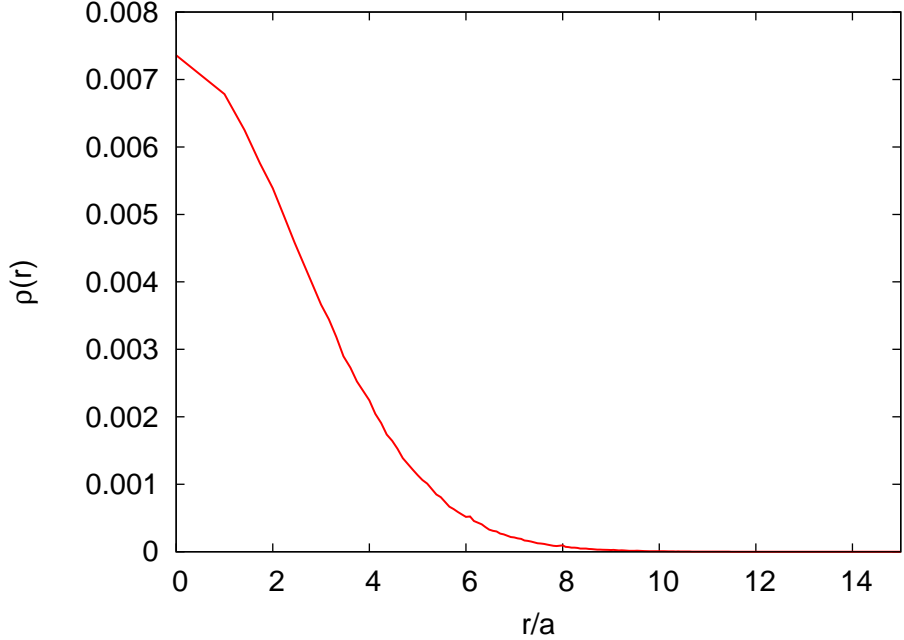}
\end{center}
\caption{
\label{plot.smear.shape}
The plot shows the profile of the smeared source vector with $w=4$ and $n=100$
on a $24^3\times 64$ configuration.  }
\end{figure}

We choose the parameters $w=4$ and $n=100$ for this work which give a smearing
size $r \sim 3$. The smeared distribution for a typical configuration is
plotted in \figref{plot.smear.shape} which is indeed close to a Gaussian shape.
This particular choice of $w$ and $n$ is aimed at increasing the overlap with
the nucleon wave function while simultaneously minimizing contamination from
different grid sites.

In addition to the smearing source, we also placed two sources on the time
slices $t=0$ and $t=32$ and calculated the inversion simultaneously. As we
shall see later in the calculation of the nucleon mass and quark scalar matrix
elements, the time window for the fitting range is between $t=6$ to $t = 14$
which is far from the two sources such that the contamination from the time
backward propagating $S_{11}(1/2^-)$ state from the second source at a distance
of $32$ time slices away is negligible. This approach nearly doubles our
statistics without computational overhead.

The local interpolating operator of the nucleon is taken to be~\cite{Wilcox:1991cq}
\begin{eqnarray}
 \chi_{\alpha} (x) &=& \epsilon^{abc}\psi_\alpha^{(u) a}(x)\psi_\beta^{(u) b} (x) (\tilde{C})_{\beta\gamma}\psi_\gamma^{(d) c}(x) \label{eq:int1} \nonumber\\
 \overline{\chi}_{\alpha'}(x)&=& -\epsilon^{a'b'c'}\overline{\psi}_ {\gamma'}^{(d) c'} (\tilde{C})_{\gamma' \beta'} \overline{\psi}_{\beta'}^{(u) b'}(x) \overline{\psi}_{\alpha'}^{(u) a'}(x),
\end{eqnarray}
where $\tilde{C} = \gamma_2 \gamma_4 \gamma_5$ in the Pauli-Sakurai gamma-matrix convention.
The color indices are denoted with latin letters and the
Dirac indices are denoted with greek letters. The nucleon correlation function
is constructed as
\begin{eqnarray}
C(y, x; \Gamma; G^{(u)}, G^{(d)}, G^{(u)}) = \langle \epsilon^{abc} \epsilon^{a^\prime b^\prime c^\prime} 
\left[ \tr\left( \Gamma G^{(u)aa^\prime}(y, x) \underline{G}^{(d)bb^\prime}(y, x) G^{(u)cc^\prime}(y, x)\right) \right. \nonumber\\
\left.+ \tr\left( \Gamma G^{(u)aa^\prime}(y, x) \right) \tr\left( \underline{G}^{(d)bb^\prime}(y, x) G^{(u)cc^\prime}(y, x)\right) \right]\rangle,
\end{eqnarray}
where the $G^{(u/d)aa^\prime}(y, x)$ stands for the $u$/$d$ quark propagator
from the site $x$ to $y$ and the color indices from $a^\prime$ to $a$. The
average is taken over the different gauge configurations and noise sources.
$\underline{G}$ is defined as $(\tilde{C}G\tilde{C}^{-1})^T$. The trace and the
transpose operations only act on Dirac indices. When the masses of the $u$ and
$d$ quarks are set to be equal, $G^{(d)}$ and $G^{(u)}$ are the same
propagator. The correlation function $C(G_1,G_2,G_3)$ is a linear functional of
the three functions $G_1, G_2$, and $G_3$.

From Eq.~\eqref{smeared_propagator}, we obtain the quark propagator as a
summation of the propagators from different grid sites with different $Z_3$
phases:
\begin{equation}
G(y,\eta^S)  = \sum_{i \in \mathcal{G}} \theta_i G_i(y),
\end{equation}
where
\begin{equation}
G_i(y) \equiv D^{-1}(y,x^\prime) S(x^\prime, x_i).
\end{equation}
The nucleon correlation function can be written as
\begin{eqnarray}
C (G,G,G) &=& \langle \hat{C}(\sum_i \theta_i G_i, \sum_j \theta_j G_j, \sum_k \theta_k G_k)\rangle \nonumber \\
           &=& \sum_{i,j,k} \langle \theta_i \theta_j \theta_k \hat{C}(G_i, G_j, G_k)\rangle \nonumber \\
           &=&  \sum_{i,j,k} \delta_{ij}\delta_{jk}\langle \hat{C}(G_i, G_j, G_k)\rangle \nonumber \\
           &\longrightarrow&  \sum_{i} \langle \hat{C}(G_i, G_i, G_i)\rangle,
\end{eqnarray}
where $\hat{C}$ denotes the correlator in a gauge
configuration. $\hat{C}(G_i, G_i, G_i)$ is the correlator of a point or a
smeared source at site $x_i$ for a gauge configuration. Thus, given a
sufficient number of noise vectors and/or gauge configurations, this
effectively increases the statistics of the correlator by the number of the
grid points as compared to that of the point source.

Since we compute the quark propagator by splitting it into the low-mode and the
high-mode pieces
\begin{eqnarray}
G& =& G^H + G^L \nonumber\\
 &=& G^H + \sum_{i}\theta_i G^L_i,
\end{eqnarray}
the nucleon correlation function can be split into contributions from the low modes
and the high modes
\begin{eqnarray}  \label{HL_correlator}
C(G,G,G) &=& C(G^H + \sum_{i}\theta_i G^L_i, G^H + \sum_{j}\theta_j G^L_j, G^H + \sum_{k}\theta_k G^L_k) 
\nonumber \\
           &=& C(G^H, G^H, G^H) + \sum_{i} C(\theta_i G^L_i, \theta_i G^L_i, \theta_i G^L_i) \nonumber \\
           &&+ \sum_{i} C(\theta_i G^L_i, G^H, G^H) + \sum_{i} C(G^H, \theta_i G^L_i, G^H) + \sum_{i} C(G^H, G^H, \theta_i G^L_i) \nonumber \\
           &&+ \sum_{i} C(\theta_i G^L_i, \theta_i G^L_i, G^H) + \sum_{i} C(\theta_i G^L_i, G^H, \theta_i G^L_i) + \sum_{i} C(G^H, \theta_i G^L_i, \theta_i G^L_i) \nonumber \\
           &&+ \sum_{i \neq j} C(\theta_i G^L_i, \theta_j G^L_j, G^H) + \sum_{i \neq j} C(\theta_i G^L_i, G^H, \theta_j G^L_j) + \sum_{i \neq j} C(G^H, \theta_i G^L_i, \theta_j G^L_j) \nonumber \\
           &&+ \sum_{i \neq j \, or \, j \neq k \, or \, k \neq i} C(\theta_i G^L_i, \theta_j G^L_j, \theta_k G^L_k),
\end{eqnarray}
where both $G^H$ and $\sum_i \theta_i G_i^L$ are noise-estimated propagators.
We denote the latter explicitly because we will modify it below to improve the signal-to-noise ratio.

We first note that the last four terms with a summation over different $Z_3$
noises, i.e. terms with $\sum_{i \neq j}$ and $\sum_{i \neq j \, or \, j \neq k
\, or \, k \neq i}$ are pure noise. They can be dropped without changing the
expectation value of the noise estimate.  For the purely high-mode
contribution, as well as the mixed terms containing two $G^H$ and a single
$G^L$, one needs to compute them with the $Z_3$ noise technique.  However, for
the correlators either involving three $G^L$ or with two $G^L$ and one $G^H$,
one can implement the LMS~\cite{Li:2010pw} to
improve the signal.  For the case where all the propagators are purely from the
low modes, we substitute the noise-estimated one [the second term on the right
side of Eq.~(\ref{HL_correlator})] with the exact expression without noise,
i.e.\ $\sum_i C(G_i^L,G_i^L,G_i^L)$ which is the sum of the low-mode
contribution to point-source correlators over the grid points. For those terms
containing only one $G^H$ and two $G^L$, we replace the noise-estimated
correlator [the sixth through eighth terms on the right side of Eq.~(\ref{HL_correlator})]
with the one where the sources of the two $G^L$ are placed at the same grid
site $i$ and multiplied with $\theta_i^2$.  This ties them with $G^H$, which
contains a $\theta$, to form the less noisy multiple-point nucleon correlator
over the grid points. (See the upper part of ~\figref{diag.grid.nucleon}.)
Since we replaced two noise-estimated $G^L$ with one noise-estimated $G^L
\otimes G^L$, this should yield a gain in the signal-to-noise ratio by
$\sqrt{V_3}$, where $V_3$ is the number of grid points.

Therefore, the nucleon correlator with LMS is  
\begin{eqnarray}
 C_{LMS}(G, G, G)      &=& C(G^H, G^H, G^H) + C(G^L, G^H, G^H) + G(G^H, G^L, G^H) + G(G^H, G^H, G^L) \nonumber \\
           &&+ \langle \sum_{i} \theta_i^2 \left[ \hat{C}(G^L_i, G^L_i, G^H) + \hat{C}(G^L_i, G^H, G^L_i) 
           + \hat{C}(G^H, G^L_i,G^L_i)  \right]\rangle \nonumber \\
           &&+  \sum_i C(G^L_i, G^L_i,G^L_i).
\end{eqnarray}

We should point out that this is different from low-mode
averaging~\cite{Neff:2001zr,Venkataraman:1997yj} in that the noise-estimated
low-mode propagators in LMS are substituted with the exact ones whenever
possible and matched with the high-mode propagators at the grid points, whereas
low-mode averaging would replace the noise-estimated low-mode part with the
average over all space-time points.

Ideally, we would compute a separate set of quark propagators from each source
point on the grid source, eliminating those gauge-dependent contributions to
the two-point function (which contribute only noise) where the propagators come
from different source points while still gaining the benefits of the extra
sources.  However, this is prohibitively expensive, as it requires additional
inversions. As mentioned previously, the grid source is chosen in order to
achieve some of the benefits of using multiple spatial sources while requiring
no additional inversions, but noise is introduced by those contributions which
mix the various source locations.  However, the purely low-mode part requires
no inversions to calculate, so it can be computed exactly without noise from
these spurious contributions.  Since this low-mode contribution dominates the
correlation function at large time separation, the use of the low modes to
compute it exactly allows for a substantial reduction in the error of the
nucleon mass.

\begin{figure}
\begin{center}
\includegraphics[width=5in]{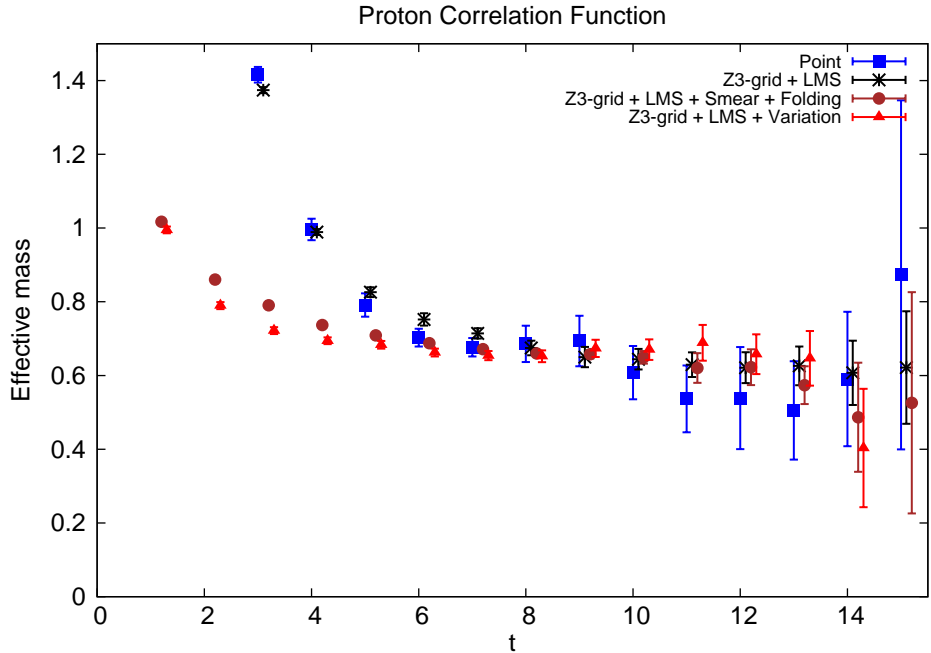}
\end{center}
\caption{
\label{plot.comp.2pt}
The comparison of proton effective masses for the point source, the $Z_3$
point-grid source, the $Z_3$ smeared-grid source and variation. The fitted
results are given in Table~\protect{\ref{Nucleon_mass}}.  }
\end{figure}

In \figref{plot.comp.2pt}, we plot the effective masses of the nucleon
correlation functions from the point source, the noise-grid source, the
noise-grid source with smearing and folding of the two correlators from sources
on two time slices at $t=0$ and $t=32$, and the variation calculation with the
point and the noise-grid source. The fitted masses are tabulated in
Table~\ref{Nucleon_mass}. They are calculated on the ensemble of 50 $2+1$
flavor DWF configurations on the $24^3 \times 64$ lattice with $a m_l=0.005$.
The overlap propagator is computed with the valence quark mass at $am \sim
0.016$ close to the light sea mass which corresponds to $m_\pi \sim 330$ MeV.
\begin{table}[!hbp]
\begin{tabular}{| c | c | c | c |  c |}
\hline
\multirow{2}{*}{} & Point &  $Z_3$-grid & $Z_3$-grid + LMS &  $Z_3$-grid + LMS \\
 &  & + LMS & + Smear + Folding & + Variation \\
\hline
 Nucleon mass (GeV) & 1.13(14) & 1.08(5) & 1.14(2) & 1.12(1) \\
\hline
\end{tabular}
\caption{\label{Nucleon_mass}
Comparison of nucleon masses from several methods.}
\end{table}
We see the fitted nucleon mass for the point source has a $12\%$ error. The
$Z_3$ noise grid source with LMS reduces the error by a factor of $2.8$ down to
$4.6\%$. Replacing the point grid with a smeared grid and putting the sources
on two time slices reduces the error further to $1.8\%$. Finally, a variational
calculation with a second nucleon interpolation field and smeared grid sources
gives an error of $1.1\%$. For this, the second nucleon interpolation field we
use is $\chi_{\alpha}' (x) = \epsilon^{abc}\gamma_5\psi_\alpha^{(u)
a}(x)\psi_\beta^{(u) b} (x) C_{\beta\gamma}\psi_\gamma^{(d) c}(x) $ and each
quark in the interpolation field can be a point-grid or a smeared-grid source.
The combination of point-grid and smeared-grid sources gives eight operators. Of these
eight, we need to consider the symmetric combination between the two $u$ quarks
so that they are spatially symmetric. This reduces the choice to six.  We
calculate all of the six combinations for each of the two interpolation fields
(which differ in their Dirac structure) and during the analysis choose four of
the twelve which have a good result for the ground state and a decent result
for the lowest excited state.  The operators of the $4\times 4$ variational
calculation are obtained by choosing $\chi$ or $\chi^\prime$ and choosing all
three quarks smeared or just one quark smeared.

It is worth emphasizing that replacing the point source by a point-grid source
with LMS, with no further improvements, reduces the error by a factor of $2.8$,
improving the statistics by a factor of 8 with no added expense.  For
comparison, we note a recent study of the all-mode-averaging (AMA)
method~\cite{Blum:2012my} on the same $24^3 \times 64$ lattices which found
that to achieve the same error on the nucleon mass with AMA on 32 smeared
sources would cost 1/4 of the computer time of a single smeared source without
AMA. This implies that AMA, given the same resources, would improve the
statistics by a factor of 4.

The variational approach using two different interpolating fields combined with
a mixed point/smeared source does reduce the variance, but not enough to
justify the increased computational cost required for the additional point-source inversion.
Overall, we find that the use of smeared $Z_3$ noise-grid sources
on two time slices is the most efficient approach; this reduces the error of
the nucleon mass by a factor of 7 compared to that of the point source while
adding only the slight overhead of LMS to the computational cost.

\section{Quark Loops}
\label{Sec:loop}

The quark loop with zero momentum on a time slice is defined as
$\sum_{\vec{x}}\tr \Gamma D^{-1}(\vec{x},\vec{x})$ where the trace is over
color and spin and $\Gamma$ is a $\gamma$ matrix. It is computed for the quark
propagator which begins and ends at the same site (or more generally a
neighboring site if a point-split current or smearing function is involved). It
is needed for meson correlators with the quark-antiquark annihilation channels
and the disconnected insertions in nucleon form-factor calculations.  Since the
quark-loop calculation involves the sum of quark propagators originating from
all lattice sites on a time slice, it is not practical to calculate it with the
usual point-source inversion. Instead, stochastic estimation is usually used
for such calculations. Taking the noise source in Eq.~(\ref{noise_source})
where the grid includes all lattice points, the loop at the position $i$ for
each gauge configuration is then estimated as
\begin{eqnarray}
L(i) &=& \langle \theta_i^* D^{-1}(i,x) \eta(x) \rangle \nonumber\\
	&=& \langle \theta_i^* \sum_{j\in \mathcal{G}} \theta_j D^{-1}(i,x) \delta_{x,j} \rangle\nonumber\\
	&\longrightarrow& D^{-1}(i,i),
\end{eqnarray}
where the average is taken over the noise. When the noise source $\eta(x)$ is
diluted in color and Dirac indices, the loop $L(i)$ is a matrix in color-spin
space. For a finite number of noises, there is noise contribution from different
sites which is not a gauge-invariant quantity and is suppressed by gauge
averaging in addition to noise averaging.  It has been shown that $Z_N$ noise
is optimal since it gives the minimum
variance~\cite{Bernardson:1993he,Dong:1993pk}. Here we shall use the $Z_4$
noise which, in addition to calculating loops, can be applied to the evaluation
of quark annihilation involving two quarks and two antiquarks in the
interpolation operators, such as in meson-meson scattering calculations.

In the nucleon three-point function involving the loops in the disconnected
insertion (DI), the quark loop including the external current needs to be
Fourier transformed to give a definite momentum transfer. Because of the
translational invariance after gauge averaging, one does not need to put the
noise on all spatial points at a given time slice. Instead, one could select an
evenly spaced grid separated by $\Delta_x$ sites in each of the spatial
directions. In this case, the Fourier transformation of $f(x)$ on a lattice
with periodic boundary conditions can project to a low-momentum $q$ in the
$x$ direction via the relation $\sum_{i\in \mathcal{G}} e^{-i q x_i} f(x_i)$.
Besides the definite momentum $q$, it involves a mixture of the next higher
momentum being $q_H= q \pm \frac{L}{\Delta_x} p_{\ell}$, where $L$ is the
spatial lattice dimension and $p_{\ell}$ is the unit of lattice momentum, i.e.\
$p_{\ell} = \frac{2\pi}{L}$.  Other higher momenta can also be mixed in, but we
will not discuss them here.  In the nucleon DI calculation, the source and sink
momenta of the nucleon propagator can have a definite momentum (the grid source
for the nucleon propagator with $\Delta_x = 6$ will have a zero-momentum source
mixed with $p = 4 p_{\ell}$). As long as these mixed momenta are taken into
account in selecting the desired momentum transfer, the mixed $q_H$ loop will
be suppressed due to momentum conservation.  Should it happen that the
contribution from the mixed high-momentum loop with $q_H$ does not vanish due
to the finite noise and gauge configurations, it will involve the intermediate
state with energy $E = \sqrt{m_N^2 + q_H^2}$.  Its contribution is
exponentially suppressed with a factor $\sim e^{-\Delta E t}$ where
$\Delta E = \sqrt{m_N^2 + q_H^2} - \sqrt{m_N^2 + q^2}$. Since $q_H > q$ and thus $\Delta E\ge 0$, the suppression
factor can be substantial for the range of $t$ where the signal of the DI calculation is to be obtained. 
For the present work, the scalar content is a forward matrix element which requires $q=0$. This makes it easier
to consider the possible $q_H$ contamination.

\begin{figure}
\begin{center}
\includegraphics[width=2.5in]{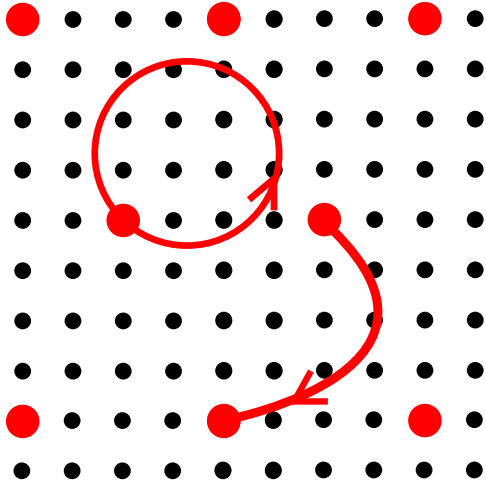}
\end{center}
\caption{
\label{diag.grid.loop}
The diagram shows the finite-noise contribution of the loops. The red dots are
the grid sources on even sites.  The closed loop shown is an example of the
signal and the open-jawed curve is an example of a noise contribution which is
suppressed by gauge and noise averaging and by the finite-distance separation
of the quark propagator.  }
\end{figure}

In our present calculation of the scalar matrix elements on the $24^3\times 64$
lattice, we shall use the grid with $(\Delta_x, \Delta_y, \Delta_z, \Delta_t) =
(4,4,4,2)$ for the high modes with odd-even dilution as well as dilution in
time. This entails the calculation of four noise propagators (two for odd-even and two
for time dilution). We show in \figref{diag.grid.loop} a cartoon of the
even-odd dilution and the time dilution.  The nearest neighbor which can give
rise to noise contribution is at a Euclidean distance of $d=\sqrt{4^2 + 2^2}
\approx 4.5$ which is reasonably far; as such, we do not do the unbiased
subtraction as has been done previously for noises which have support on all
lattice points~\cite{Thron:1997iy,Dong:1995ec}, where the nearest neighbor is a
single lattice site away. Should there be noise contribution from the higher
mixed momentum $q_H$ despite suppression due to momentum conservation as
discussed above, it will be further suppressed by $e^{-\Delta E t}$. With $q_H
= 2 \pi/4, q = 0$ and $m_N \sim 0.66$, $\Delta E$ is 1.04. As a result, the
suppression factor for $t \ge 6$ where the strangeness matrix element will be
extracted is less than $e^{-6.24}= 0.002$ which renders this noise contribution negligible.

To further improve the loop calculation, we use low-mode averaging (LMA) for
the low-mode contribution by summing over the spatial volume on a time slice,
while the high-mode contribution is calculated on the grid as described above.
\begin{equation}
L = \sum_i L_L(i) + \sum_j \frac{L^3}{\Delta_x \Delta_y \Delta_z} L_H(j),
\end{equation}
where $L_H$ is scaled from the sum of the grid points to that of the full space
volume to match with $L_L(i)$ from LMA.

\begin{figure}
\begin{center}
\includegraphics[width=6in]{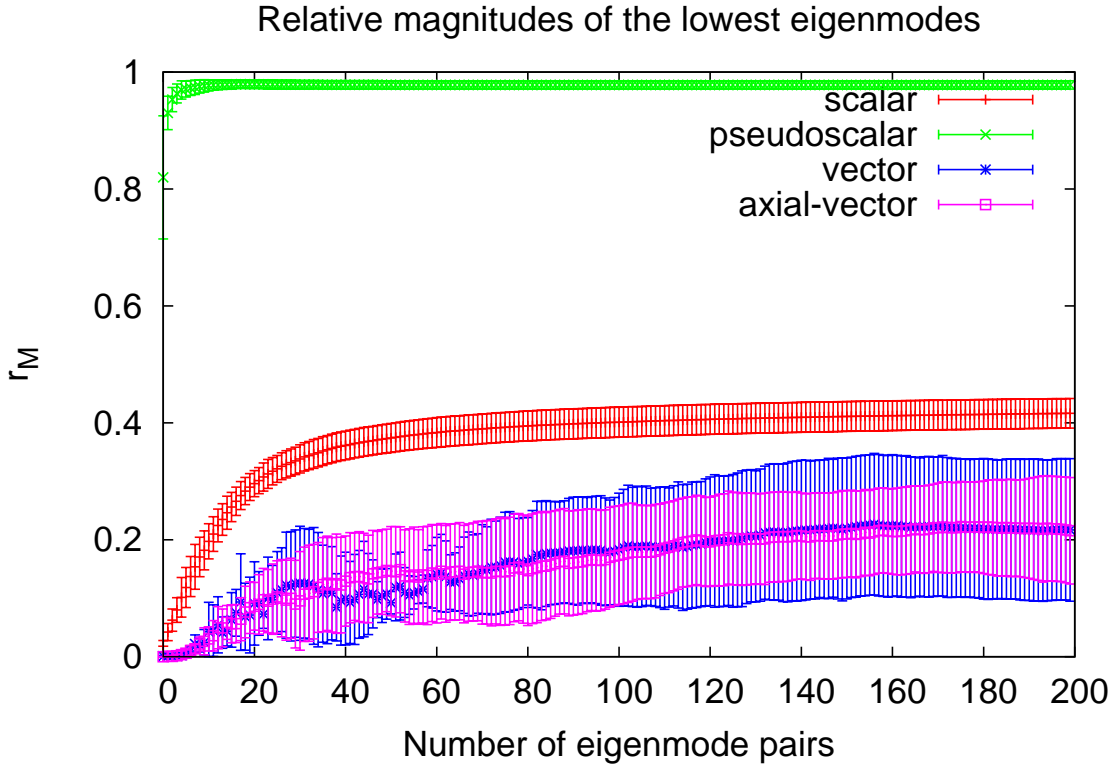}
\end{center}
\caption{
\label{plot.sat.loop.neig}
The plot shows the relative low-mode magnitude $r_M$ defined in
Eq.~(\protect\ref{sat_loop}) as a function of the accumulated pairs of
eigenmodes plus the zero modes. The corresponding pion mass is about $330$ MeV
for both the quark in the loop and the light sea-quark. The error bars denote
the standard deviation from five gauge configurations.  }
\end{figure}  

It is interesting to find out the individual contributions to the quark loops
from the low modes and the high modes.  To this end, we measure the following
quantity to gauge the magnitude of the low-mode contribution relative to the
sum of those of the low modes and the high modes,
\begin{equation} \label{sat_loop}
r_M= \frac{ \left| L^{low} \right| }{\left| L^{low} \right| + \left| L^{high} \right|}.
\end{equation}

The norms of the low- and the high-mode parts of the loops are used to avoid
the singular situation when the full loop is near zero, especially for the
vector and axial-vector channels, where the vacuum expectation values of the
loops are zero. The ratio is taken before averaging over the time slices so as
to be more relevant to the DI calculation where the loop is to be correlated
with the nucleon propagator in a limited time range. We used five configurations
in this study.

\begin{figure}
\begin{center}
\includegraphics[width=6in]{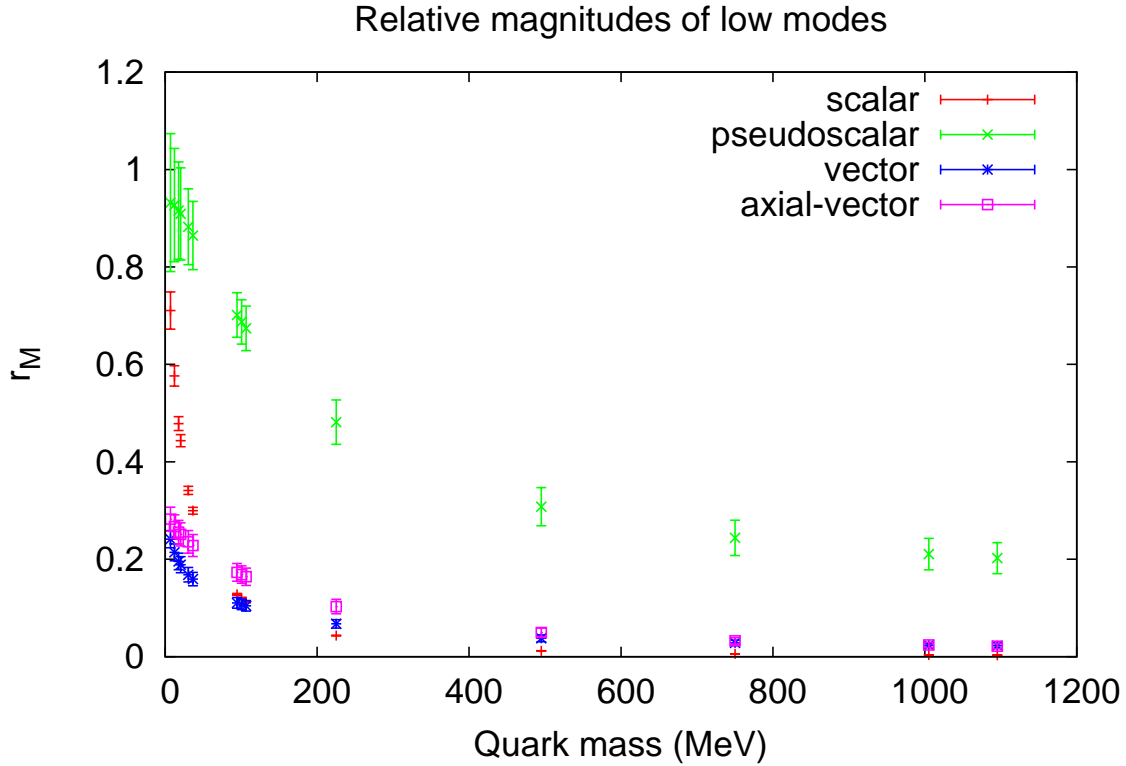}
\end{center}
\caption{
\label{plot.sat.loop.mass}
The plot shows the relative low-mode magnitude $r_M$ with different bare quark
masses for the loop.  The error bars denote the standard deviation over gauge
configurations.  This study is carried out on 100 configurations.  }
\end{figure}

The relative magnitudes of the low modes, defined as $r_M$ in
Eq.~(\ref{sat_loop}), in terms of the accumulated pairs of low eigenmodes in
addition to the zero modes are shown in \figref{plot.sat.loop.neig} for
different currents. This is for the case of a quark mass which corresponds to a
pion mass of $\sim 330$ MeV.  We see that $r_M$ of the pseudoscalar loop is
dominated by a few lowest eigenmodes, and all the higher modes contribute only
a few percent. This is not surprising, since the space-time integral of the
pseudoscalar density times the quark mass is just the topological charge of the
configuration which is totally determined by the zero modes.  For the scalar
loop, both the $L^{\mathrm{low}}$ and $L^{\mathrm{high}}$ are positive on all
time slices and for all configurations. Therefore, the relative magnitude $r_M$
is the relative low-mode contribution. In this case, $\sim 40\%$ of the scalar
loop is contributed by the lowest 50 pairs of eigenmodes. Even at this low
quark mass, most of the contribution comes from the high modes. In the vector
and axial-vector channels, the relative low-mode magnitude is small with a
large variance.

To study the quark-mass dependence, we show $r_M$ in
\figref{plot.sat.loop.mass} with 200 pairs of the lowest eigenmodes plus zero
modes as a function of the quark mass.  The relative magnitude of the low modes
in the pseudoscalar channel dominates in the light quark region and decreases
with quark mass. The $r_M$ for the scalar, vector, and axial loops decrease
with quark mass faster than that of the pseudoscalar loop and become very small
for $m > 500$ MeV.  This hints that for the charmness in the nucleon, the high
modes may play a substantial role.

\section{Disconnected Insertion of The Three-Point Correlation Function}
\label{Sec:3pt}

\begin{figure}
\begin{center}
\includegraphics[width=3in]{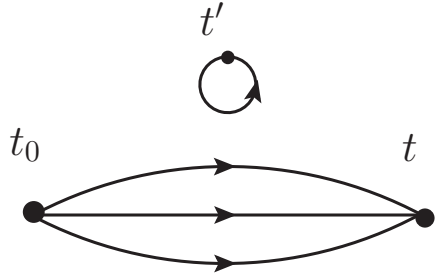}
\end{center}
\caption{
\label{diag.plateau}
The sketch illustrates the disconnected insertion of the three-point function.
}
\end{figure}

In order to extract the strangeness and charmness contributions to the nucleon
given in Eq.~\eqref{f_T}, we compute the ratio of the disconnected three-point
function to the two-point function, defined as
\begin{equation}
R(t, t^\prime, t_0) = \frac{< \chi_{N}(t) \,\bar{q}q(t^\prime)\, \bar{\chi}_{N}(t_0) > 
- < \chi_N(t) \bar{\chi}_N(t_0) > < \bar{q}q(t^\prime)>}{< \chi_N(t) \bar{\chi}_N(t_0) >}.
\end{equation}
The matrix element for the quark content in the nucleon can be extracted from
this ratio,
\begin{eqnarray}
f_{T_{q}} &=& \frac{m_q}{m_N} \left< N | \bar{q} q | N \right> \nonumber \\
	&=& \frac{m_q}{m_N} \lim_{\overset{t^\prime-t_0\rightarrow\infty}{t-t^\prime\rightarrow\infty}} 
	R(t, t^\prime, t_0).
\label{quarkcontent}
\end{eqnarray}
For the disconnected insertion, one can write the above ratio in terms of the
nucleon propagator and the quark loop as in Sec.~\ref{Sec:loop},
\begin{equation}
R(t, t^\prime, t_0) = \frac{\left< \hat{C}(t, t_0) \left( L(t^\prime) - \left< L(t^\prime) \right> \right) \right> }
{\left< \hat{C}(t, t_0) \right>}.
\end{equation}
This approach is illustrated in \figref{diag.plateau}.

\begin{figure}
\begin{center}
\includegraphics[width=5in]{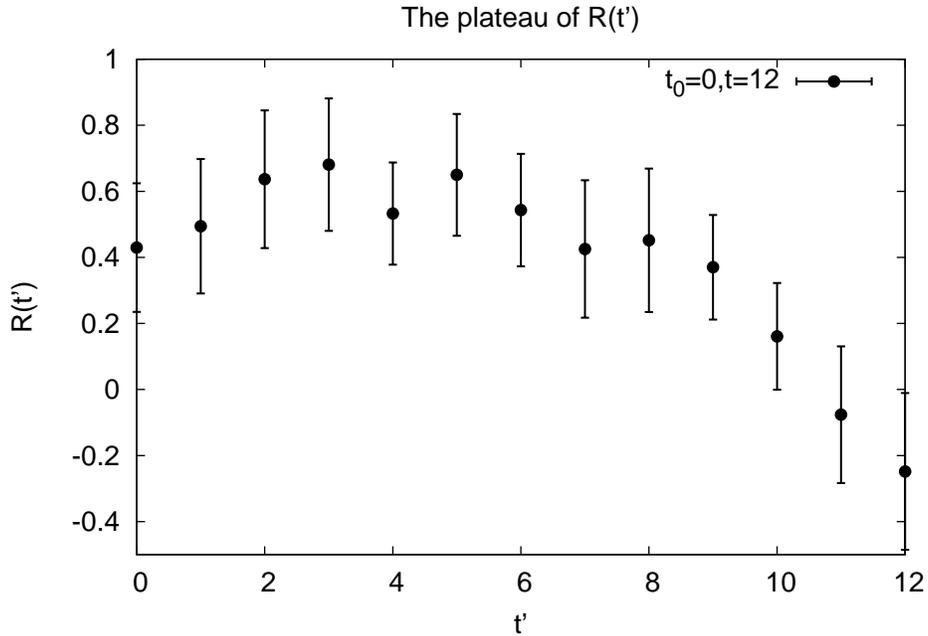}
\end{center}
\caption{
\label{plot.plateau}
The plot shows the behavior of $R(t, t^\prime, t_0)$ as a function of
$t^\prime$ for fixed $t_0$ and $t$ at 0 and 12.  The expected plateau in the
middle of the time window is rather noisy.  }
\end{figure}

We show in \figref{plot.plateau} $R(t,t',t_0)$ as a function of $t'$ from
$t_0=0$ to $t=12$ for the case where the quark mass in the loop is $0.063$,
which corresponds to the strange quark mass, and the quark mass in the nucleon
propagator is $0.016$, which corresponds to $m_{\pi} = 330$ MeV.  This
calculation is done on 176 configurations.  One can see that the error bars are
quite large, and it appears to be difficult to determine a plateau region from
which to extract the matrix element.

\begin{figure}[hb]
\vspace*{0.5cm}
\subfigure[]
{\includegraphics[height=1.5in]{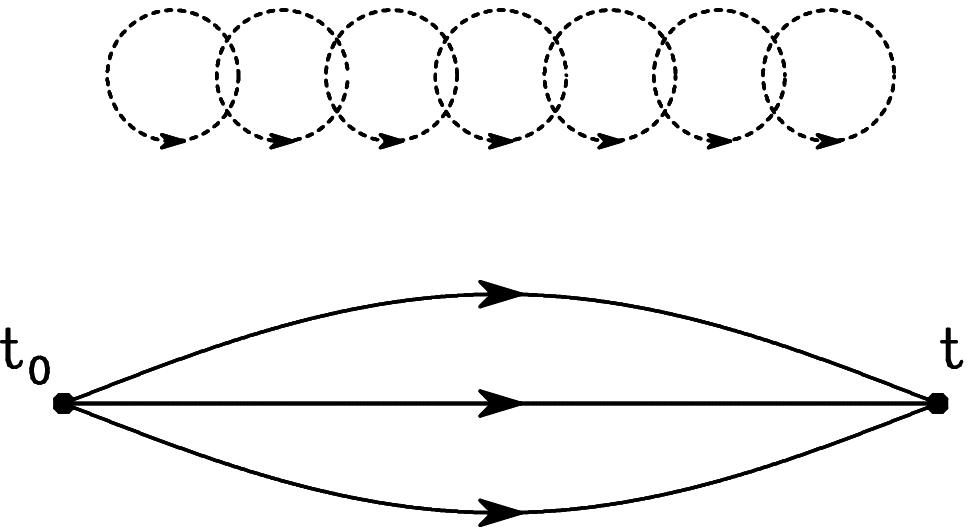}}
\hfill
\subfigure[]
{\includegraphics[height=1.5in]{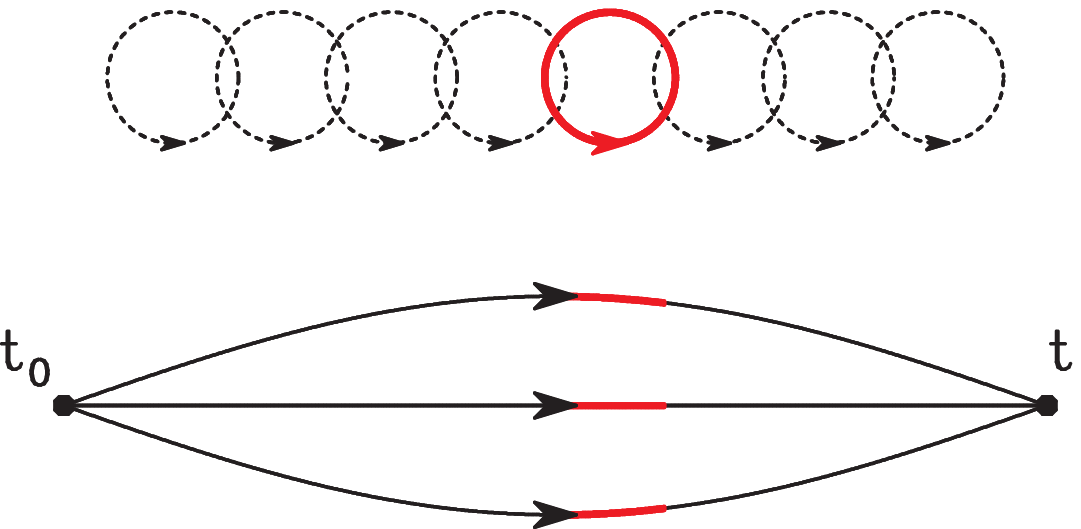}}
\vspace*{-0.5cm}
\caption{
\label{diag.sum}
The cartoon shows the summed ratio method for the disconnected three-point function.
The red part is the additional contribution to the sum when the propagator length is increased, which is within the plateau region.
}
\end{figure}

One can improve the statistics by using the summed ratio
method~\cite{Maiani:1987by}. As illustrated in \figref{diag.sum},
$R(t,t^\prime, t_0)$ is summed over $t'$ between $t_0+1$ and $t-1$ inclusive
\begin{equation}
R^\prime(t, t_0) = \sum_{t^\prime=t_0+1}^{t-1} R(t, t^\prime, t_0).
\end{equation}
Previous calculations of the quark momentum fraction $\left<x\right>$ in the
DI~\cite{Deka:2008xr}, have studied various choices for the domain of
$t^\prime$.  They found that choosing $t^\prime$ between $t_0+1$ and $t-1$
inclusive produced less noise than other choices.  For this reason, we have
chosen to adopt the same domain in our calculations.

The ratio method is a means to incorporate information from multiple values of
$t-t_0$ and $t^\prime$, thus reducing the statistical error, without needing to
explicitly consider the points at which $t^\prime$ might suffer from excited-state
pollution because they are too close to $t_0$ and $t$. The method relies on the
fact that when the value of $R$ is summed from source to sink, the contribution
from these contaminated points near the end does not depend on the distance
from source to sink so long as that distance is large enough. When the
propagator length is increased, the additional contribution to the sum comes
from a single point in the center, within the plateau region. Thus, by
examining how much the summed ratio increases when the propagator length is
increased, we can get an estimate for the plateau value of $R$ that
incorporates information from multiple values of $t^\prime$ and $t$.

It was shown~\cite{Maiani:1987by,Deka:2008xr} that the ratio $R'(t,t_0)$ has a
linear behavior in a region where $t$ is large enough to have a plateau in
$R(t,t',t_0)$ and thus
\begin{equation}
R^\prime(t, t_0) {}_{\stackrel{\longrightarrow}{t \gg t_0}} const + t \left< N | \bar{q} q | N \right>.
\label{fit}
\end{equation}
We fit the value of the summed ratio $R'$ to constant-plus-linear over the
range in which $t$ is large enough for $R$ to have a plateau; the slope in $R'$
is thus a measurement of the plateau value of $R$, i.e.\ the matrix element,
from which the quark content $f_{T_{q}}$ is determined using
Eq.~\eqref{quarkcontent}.  We should mention that the excited-state
contamination in the sum method is found to be $O(e^{-\Delta_E
t})$~\cite{Deka:2008xr,Capitani:2012gj} where $\Delta_E$ is the energy gap
between the nucleon ground state and the lowest excited state and $t$ is the
sink time of the nucleon propagator (we have taken the source time at
$t_0=0$). On the other hand, the contamination of the excited state in the
plateau method is $O(e^{-\Delta_E t_p}+(e^{-\Delta_E (t_s - t_p)}))$
where $t_p$ is the time when the plateau appears and $t_s$ is the fixed sink
time of the nucleon propagator.  Since in practice $t \sim t_s > t_p$ and $t
\sim t_s > (t_s - t_p)$, the excited-state contamination is less than that of
the plateau method and this has been demonstrated in the calculation of the
isovector $g_A$~\cite{Capitani:2012gj}.

\section{Numerical Results}
\label{Sec:results}

\begin{figure}[hbt]
\begin{center}
\subfigure[]
{\includegraphics[width=0.49\textwidth]{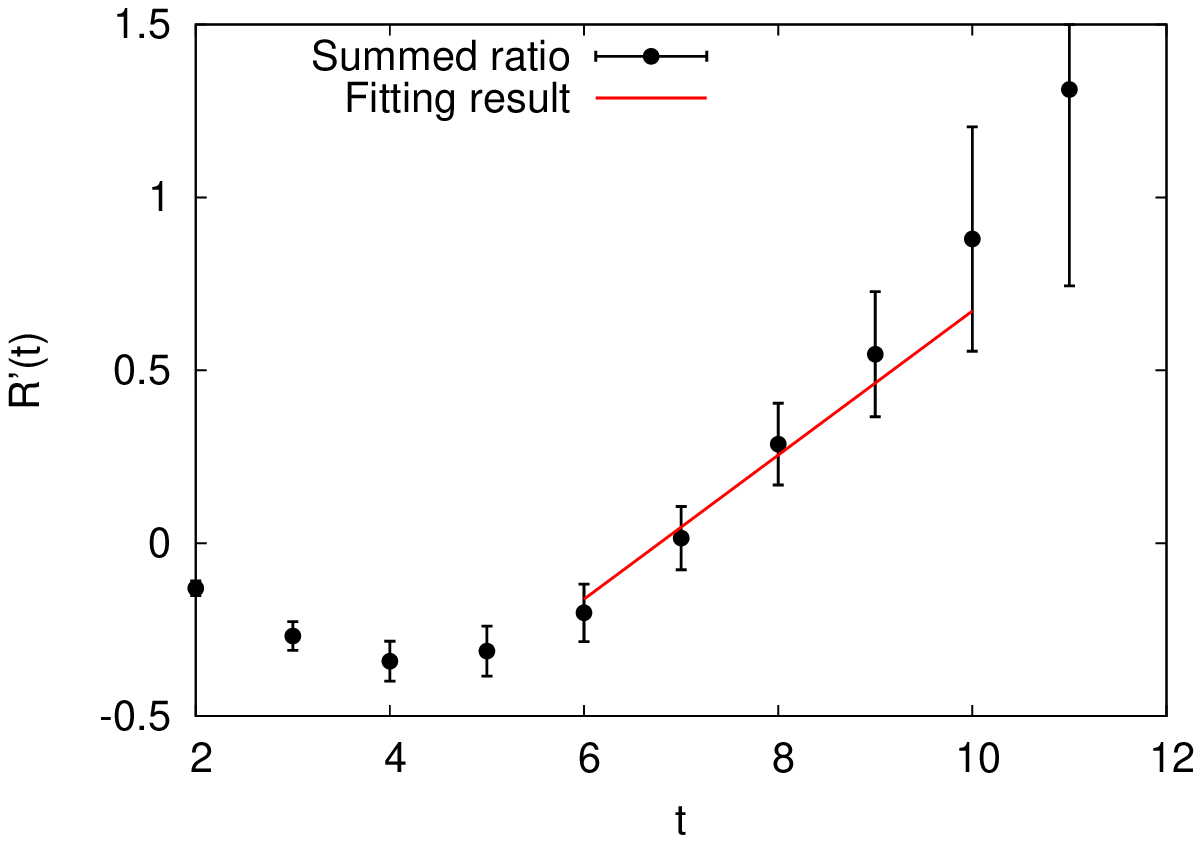}\label{plot.sum.strange}}
\hfill
\subfigure[]
{\includegraphics[width=0.49\textwidth]{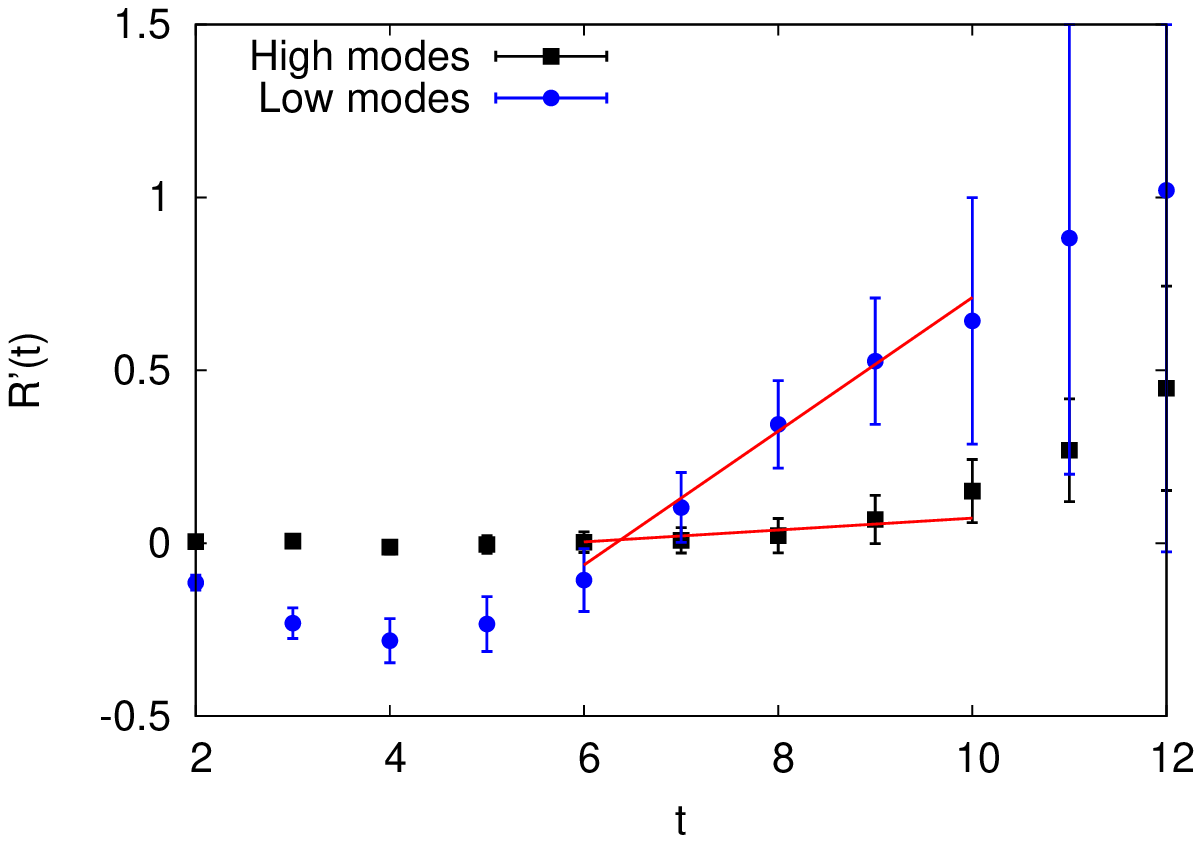}\label{plot.sum.strange.hl}}
\end{center}
\caption{(a) The plot for the summed ratio of the strangeness.
The corresponding pion mass is $330$ MeV. (b) The separate
contributions from the low modes and the high modes.}
\end{figure}

This work is carried out using valence overlap fermions on 176 configurations
with $2+1$ flavors of sea domain-wall fermions.  The lattice size is
$24^3\times 64$ with the inverse lattice spacing \mbox{$1/a = 1.73(3)$ GeV.}
The $u$/$d$ quark mass in the DWF sea is $0.005$ which corresponds to a pion
mass of $331$ MeV.  The sea strange-quark mass is $0.04$. This pion mass is
matched for the overlap fermion with a valence $u$/$d$ quark mass of
0.016. With the help of the multimass algorithm~\cite{Jegerlehner:1996pm}, we
compute a set of 26 quark masses for the nucleon and 10 for the loops with an
overhead of about 8\%~\cite{Ying:1996je} of the cost of inverting the lowest
quark mass.  The summed ratio of the strangeness is shown in
\figref{plot.sum.strange} for $am_s = 0.063$ and $am_{ud} = 0.016$.

The nucleon propagator in the DI of the three-point function is calculated with
a smeared grid source at $t=0$ and $t=32$. The time-forward propagator with
positive-parity projection and the time-backward propagator with
negative-parity projection are averaged. We see that a linear slope develops
after $t=6$ where the nucleon starts to emerge in the nucleon correlator and it
is in the middle of the plateau from~\figref{plot.plateau}.

To see the contributions to $R'(t,t_0)$ and the respective slopes from the low
modes and the high modes in the loop, we plot them separately in
Fig.~\ref{plot.sum.strange.hl}. It is interesting to observe that practically
all the contributions to $R'(t,t_0)$ are coming from the low modes despite the
fact that the low modes saturate only $\sim 15\%$ of the strange quark
loop. The high-mode contribution is quite small.  Since the low-mode
contribution is exact, it only has variance from the gauge ensemble. The
variance of the high-mode contribution comes from both the noise and gauge
ensembles. Since its contribution is small, there is no need to improve this
part of the estimate with more noise vectors.

\begin{figure}
\begin{center}
\includegraphics[width=5in]{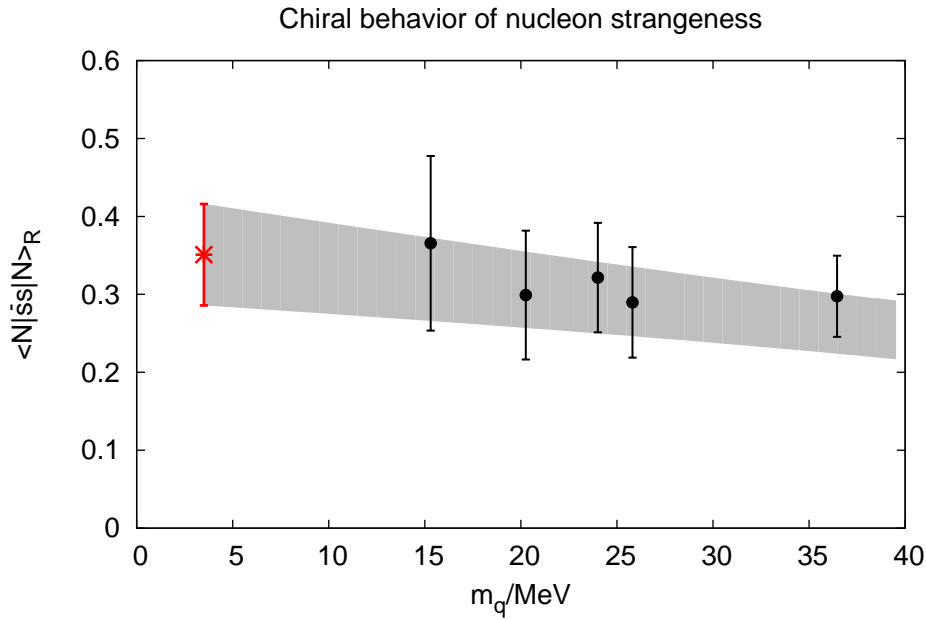}
\end{center}
\caption{
\label{plot.ch.s}
The dependence of the renormalized strangeness content on the renormalized
$u$/$d$ quark mass in the nucleon propagator.  The round dots show the data
points with $a m_s=0.063$ for reference and the error band shows the global
fitting results with physical strange mass. }
\end{figure}

The renormalized strangeness content reads
\begin{equation}
\langle N|\bar{s}s|N\rangle_R=Z_S^{\overline{MS}}(2 GeV)\,\langle N|\bar{s}s|N\rangle,
\end{equation}
where the nonperturbative renormalization constant $Z_S^{\overline{MS}}(2 GeV)
= 1.121(6)$ (stat) has been calculated in the regularization-independent momentum-subtraction renormalization scheme~\cite{zfliu13} with
input of $Z_A$ from the chiral Ward identity. It is plotted in
\figref{plot.ch.s} as a function of the valence $u/d$ quark mass.  The results
are obtained from fitting the range from $t=7$ to 14 in each quark-mass case.
The lowest quark mass gives $m_{\pi} = 250$ MeV.

To obtain a value at the physical point we must perform a chiral extrapolation.
The mixed-action formula can be derived from the partially quenched expression in Ref.~\cite{Chen:2002bz},
using the simple conversion of partially quenched to mixed-action formulas described in Refs.~\cite{Chen:2006wf,Chen:2007ug}.
The next-to-leading-order mixed-action extrapolation formula is~\cite{walker-loudnote}
\begin{eqnarray}
\left< N | \bar{s} s | N \right> &=& {\left< N | \bar{s} s | N \right>}^{LO} + \frac{2 g^2_{\Delta N}}{(4\pi f_\pi)^2} \left(  {\left< N | \bar{s} s | N \right>}^{LO} -  {\left< \Delta | \bar{s} s | \Delta \right>}^{LO} \right) \left( J(m_\pi, \Delta, \mu) + J(\tilde{m}_{ju}, \Delta, \mu) \right) \nonumber\\
		&& + E_s(\mu) \frac{m_\pi^2}{(4\pi f_\pi)^2} + E_s^{PQ}(\mu) \frac{m_{ju}^2-m_\pi^2}{(4\pi f_\pi)^2} + E_s^a(\mu) \frac{a^2\Delta_{mix}}{(4\pi f_\pi)^2},
\label{eq:chiral0}
\end{eqnarray}
where $\left< N,\Delta | \bar{s} s | N \Delta\right>^{LO}$ are the leading-order contributions to the strange
matrix element in the chiral limit and 
the nonanalytic chiral loop function is~\cite{Tiburzi:2005na}
\begin{equation}
J(m, \Delta, \mu) = 2\Delta \sqrt{\Delta^2-m^2+i\epsilon}\log\left(\frac{\Delta-\sqrt{\Delta^2-m^2+i\epsilon}}{\Delta+\sqrt{\Delta^2-m^2+i\epsilon}}\right) + m^2\log\left(\frac{m^2}{\mu^2}\right) + 2 \Delta^2 \log\left( \frac{4\Delta^2}{m^2} \right),
\end{equation}
where $\Delta$ is the mass difference of the $\Delta$ baryon and the nucleon.

In the $SU(2)$ case without an explicit delta degree of freedom, Eq.~\eqref{eq:chiral0} can be simplified to
\begin{equation}
\left< N | \bar{s} s | N \right> = {\left< N | \bar{s} s | N \right>}^{LO} + E_{s,\cancel\Delta}(\mu) \frac{m_\pi^2}{(4\pi f_\pi)^2} + E_{s,\cancel\Delta}^{PQ}(\mu) \frac{m_{ju}^2-m_\pi^2}{(4\pi f_\pi)^2} + E_{s,\cancel\Delta}^a(\mu) \frac{a^2\Delta_{mix}}{(4\pi f_\pi)^2}.
\label{eq:chiral1}
\end{equation}
The only mixed action low-energy
constant $\Delta_{mix}$~\cite{Chen:2006wf} between the valence overlap fermion
and sea domain-wall fermions is found to be small~\cite{Lujan:2012wg} --- it
shifts the pion mass at $300$ MeV by a mere 16 MeV.  We can safely neglect the
mixed-action effects since they are quite small compared to the statistical
error.

Combining Eq.~\eqref{fit} and Eq.~\eqref{eq:chiral1}, we obtain a
global fit using different time slices and quark masses. The fitting model for
strangeness reads
\begin{equation}
\label{fittingmodel}
R^\prime(t, t_0) {}_{\stackrel{\longrightarrow}{t \gg t_0}} \left[ \left< N | \bar{s} s | N \right> + A(m_l-m_l^0) + B(m_s-m_s^0) \right] t + C_{m_l,m_s},
\end{equation}
where $m_l^0$ and $m_s^0$ are the physical quark masses corresponding to the
correct $\pi$ and $K$ masses, and $C_{m_l,m_s}$ is a set of constants.  

\begin{figure}
\begin{center}
\includegraphics[width=5in]{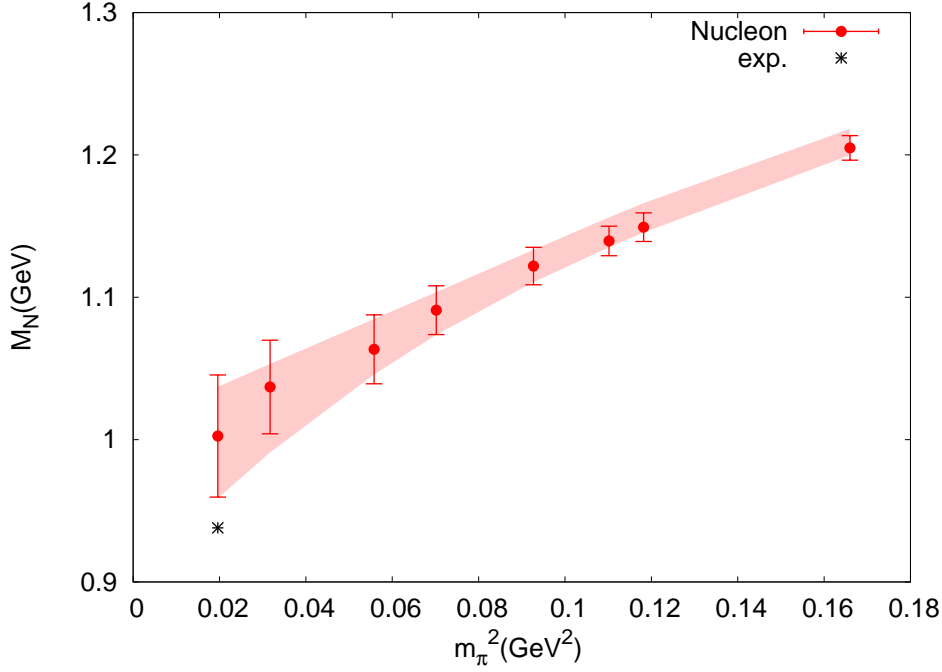}
\end{center}
\caption{
\label{nucleonextra}
The chiral extrapolation of the nucleon mass.}
\end{figure}

\begin{figure}
\begin{center}
\includegraphics[width=5in]{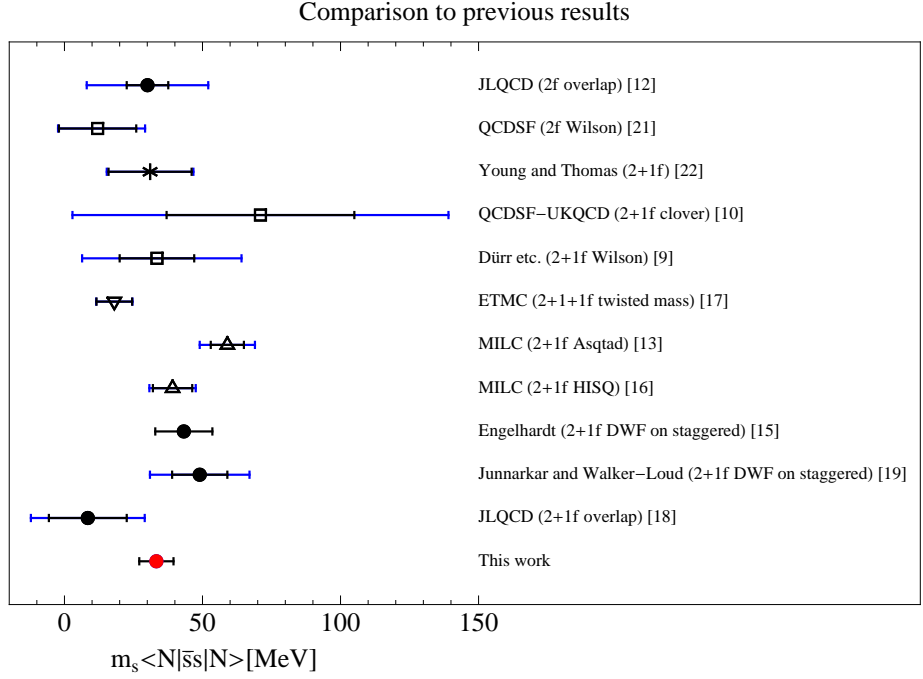}
\end{center}
\caption{
\label{plot.compare.strangeness}
A comparison of the result of our calculation of the strangeness $\sigma$ term
$\sigma_s = m_s \langle N |\bar{s}s|N\rangle$ with those of other groups.
The statistical errors are denoted by black error bars and the total errors are denoted by blue error bars.
}
\end{figure}

After the chiral extrapolation in the valence quark mass and the interpolation
in the loop quark mass, we get the renormalized strangeness matrix element
$\langle N|\bar{s}s|N\rangle_R = 0.341(63)$ and $m_s \langle N|\bar{s}s|N\rangle =33.3(6.2)$ MeV
with a fitting range from $t=7$ to 14
and $a m_s^0=0.063$.
We also calculate the nucleon mass and extrapolate it to the chiral limit with
$m_N(m_\pi) = m_N(0) + C_1 m_\pi^2 + C_2 m_\pi^3$ as shown in \figref{nucleonextra}
and we obtain $m_N(0) = 0.998(39)$ GeV at the physical pion mass.
Using this number, we obtain
$f_{T_{s}} = 0.0334(62)$.
We plot the recent results of $m_s \langle N|\bar{s}s|N\rangle$ in
\figref{plot.compare.strangeness} from calculations with dynamical fermions
with $N_f = 2$ and $N_f = 2+1$ (the ETM collaboration's calculation is with $N_f = 2+1+1$).  We see
that our result has a small statistical error and its $5 \sigma$ relative error is
comparable to those of
Refs.~\cite{Toussaint:2009pz,Engelhardt:2012gd,Freeman:2012ry}.

\begin{figure}
\subfigure[]
{\includegraphics[width=0.49\textwidth]{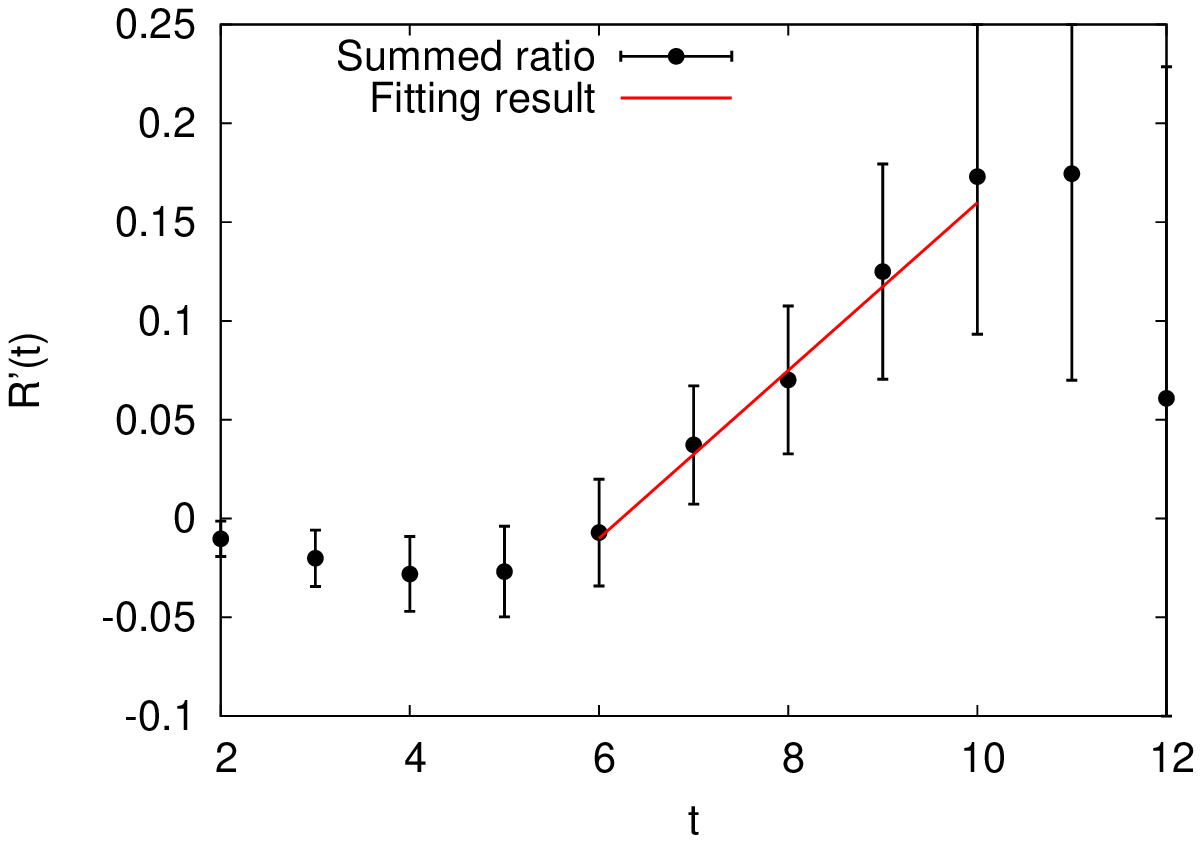}\label{plot.sum.charm}}
\hfill
\subfigure[]
{\includegraphics[width=0.49\textwidth]{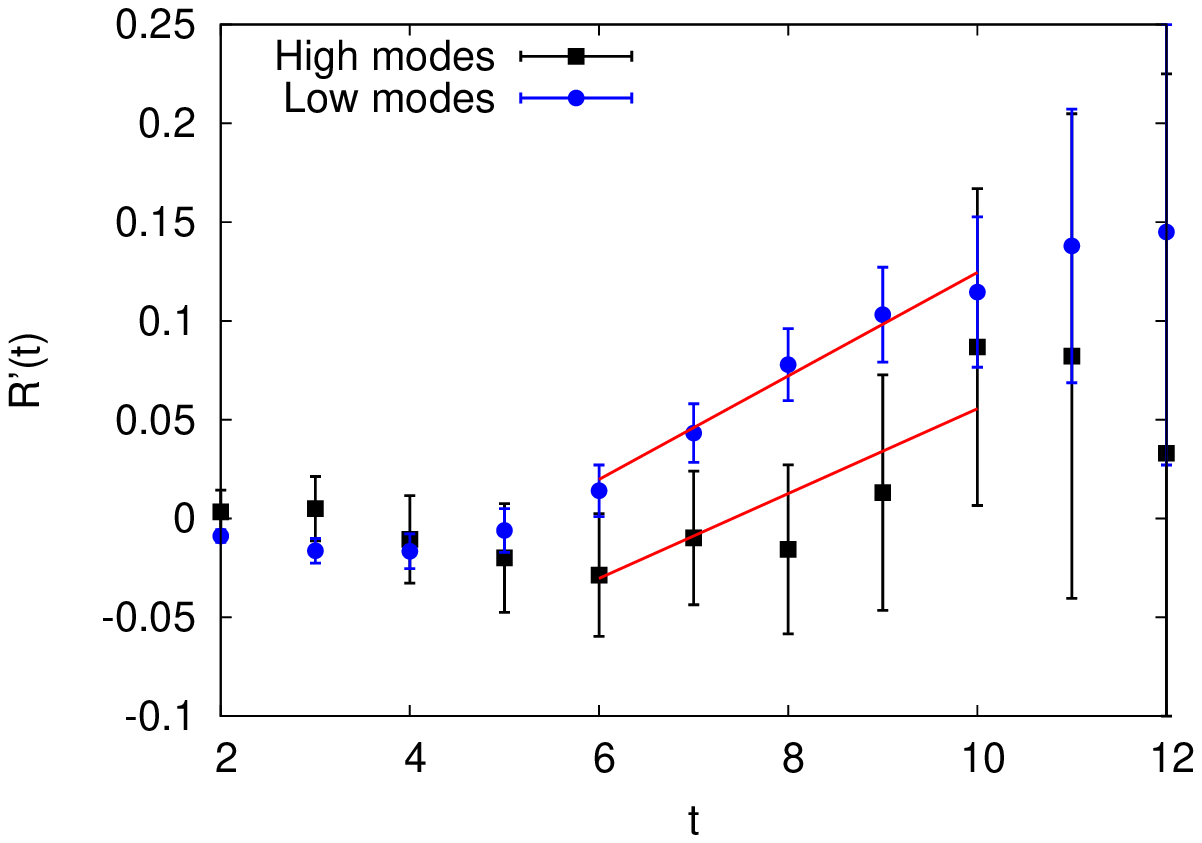}\label{plot.sum.charm.hl}}
\caption{The summed ratio of the charmness is plotted in the left panel. The separate high-mode and low-mode parts are plotted in the right panel. 
These results are computed with a valence light quark mass which corresponds to $m_\pi = 330$ MeV.}
\end{figure}

Similarly, we can compute the charmness content $m_c \langle N|\bar{c}c|N\rangle$ using the same method. 
The summed ratio of the charmness is shown in \figref{plot.sum.charm} and the low/high separation is
given in \figref{plot.sum.charm.hl}. We see that the low-mode part still plays an important role, but
in contrast to the strangeness case, the high-mode contribution is no longer small. 

\begin{figure}
\begin{center}
\includegraphics[width=5in]{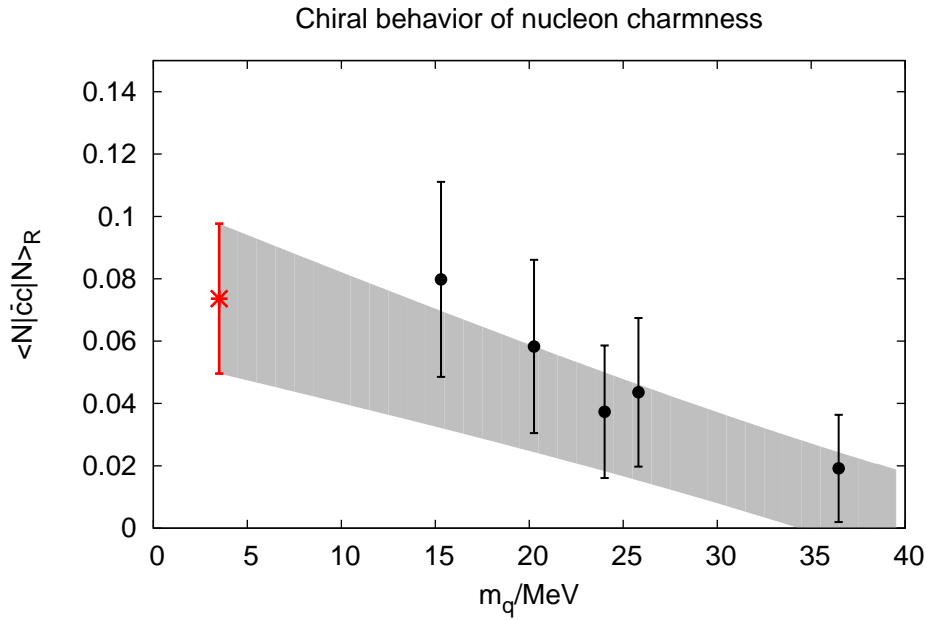}
\end{center}
\caption{
\label{plot.ch.c}
The dependence of the charmness content on the $u$/$d$ quark mass in the nucleon propagator.
The round dots show the data points with $a m_c=0.67$ for reference
and the error band shows the global fitting results with physical charm mass.
}
\end{figure}

The chiral extrapolation is shown in \figref{plot.ch.c}.
We take $a m_c^0 = 0.67$ which corresponds to the physical charm mass in the global analysis of the charmonium spectrum with three sea masses and two lattice spacings~\cite{ybyang13}.
We find the renormalized charmness content $\langle N|\bar{c}c|N\rangle_R= 0.072(23)$ and
$f_{T_{c}} = 0.094(31)$ after the chiral extrapolation and charm-quark-mass interpolation.
This number is consistent with the result from the MILC collaboration at $\langle
N|\bar{c}c|N\rangle=0.056(27)$\cite{Freeman:2012ry}, but with a smaller
relative error at 33\%.
It is interesting to note that $m_c\langle N|\bar{c}c|N\rangle$ at $94(31)$ MeV
is larger than $m_s\langle N|\bar{s}s|N\rangle = 33.3(6.2)$ MeV and agrees with
the prediction of 70 MeV in Eq.~(\ref{heavy_quark}) from the heavy-quark
expansion~\cite{Shifman:1978zn}.

\begin{figure}
\begin{center}
\includegraphics[width=5in]{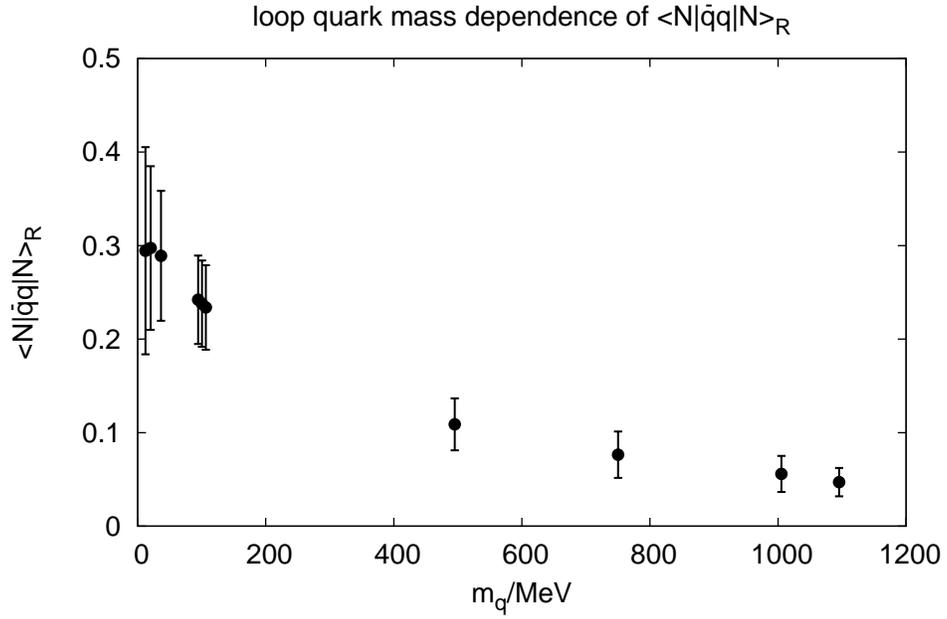}
\end{center}
\caption{
\label{plot.heavy}
The dependence of $\langle N|\bar{q}q|N\rangle_R$ on the loop quark mass, with
the mass of the nucleon valence quarks fixed at a value corresponding to
$m_{\pi} = 330$ MeV.}
\end{figure}

\begin{figure}
\begin{center}
\includegraphics[width=5in]{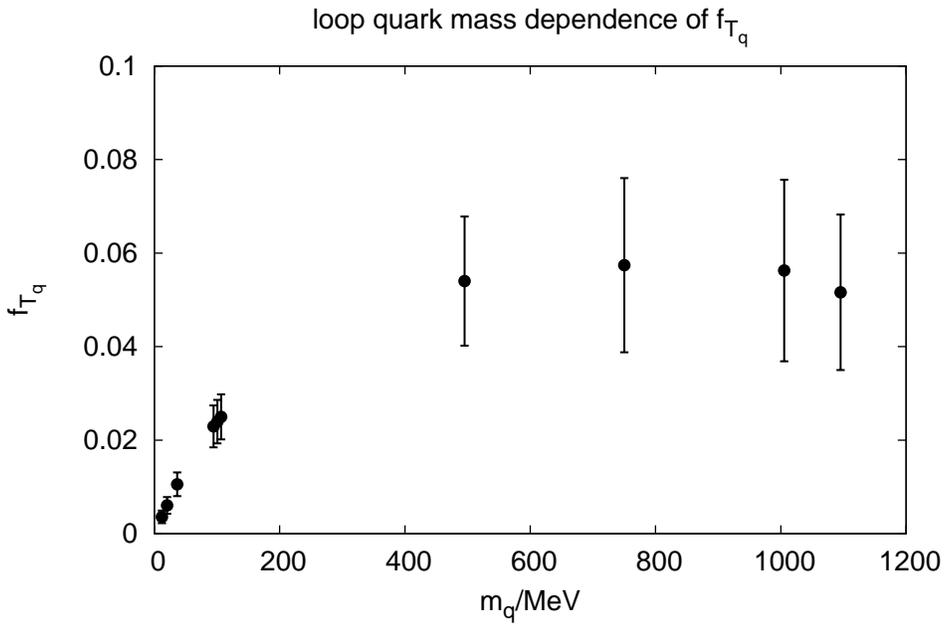}
\end{center}
\caption{
\label{plot.heavy.ftq}
The same as the last figure for the dependence of $f_{T_{q,Q}}$ of the nucleon
on the quark mass.  }
\end{figure}

By varying the quark mass in the loop with a fixed light quark mass in the
nucleon, we can check the quark-mass dependence of $f_{T_q}$ and $f_{T_Q}$.
Figures \ref{plot.heavy} and \ref{plot.heavy.ftq} display the quark-mass
dependence of $\langle N|\bar{q}q|N\rangle_R$ and $f_{T_{q,Q}}$ respectively as
a function of the quark mass. We see that the matrix element seems to go down
as $1/m_q$ at reasonably large $m_q$ and $f_{T_q}$ appears to be flat beyond $m_q
\sim 500$MeV.  This behavior will be checked with higher precision on the $32^3
\times 64$ lattice with $a^{-1} = 2.35$ GeV which can accommodate heavier quark
masses than on the present $24^3 \times 64$ lattice.

\section{Conclusion}
\label{Sec:conclusion}

In this work, we have computed both the strangeness and the charmness content
of the nucleon with overlap valence fermions on $2+1$-flavor dynamical DWF
configurations on the $24^3\times 64$ lattice with $a^{-1} = 1.73$ GeV and a
sea pion mass of 331 MeV.  We have employed a smeared $Z_3$ noise-grid source
with low-mode substitution to calculate the nucleon two-point functions, which
reduces the error on the nucleon mass by a factor of 7 compared to the
calculation with a point source.  For the loop part of the three-point
disconnected insertion calculation, we used low-mode averaging to compute the
low-mode part exactly and used the $Z_4$ noise on a grid to estimate the high-mode part.
It turns out that the high-mode part contributes negligibly to the
strangeness content.  With the highly improved nucleon propagators and the
quark loops, we extrapolated to the physical pion mass and obtained the precise
value $f_{T_{s}} = 0.0334(62)$ with a better than 5 $\sigma$ signal. The
statistical error is quite small compared to those of other lattice calculations.  The renormalized
matrix element is $\langle N|\bar{s}s|N\rangle_R = 0.341(63)$ and the strange-quark sigma term $\sigma_s$ is $m_s \langle N|\bar{s}s|N\rangle =33.3(6.2)$ MeV.
Similarly, we obtain $f_{T_{c}} = 0.094(31)$ with a 3 $\sigma$ signal which is
the first time such a result has been obtained for the charm beyond a precision of
two sigma.  The renormalized matrix element is $\langle N|\bar{c}c|N\rangle_R =
0.072(23)$ and the charm-quark sigma term is $m_c \langle N|\bar{c}c|N\rangle
=94(31)$ MeV which is consistent with the prediction based on the heavy quark
expansion~\cite{Shifman:1978zn,Kryjevski:2003mh}.

Even though our present work has a high precision, a more meaningful
comparison would be with the number of inversions one has to do in order to
achieve the same precision.  To this end, we shall compare our results with the
calculations by Engelhardt~\cite{Engelhardt:2012gd} and
JLQCD~\cite{Oksuzian:2012rzb}.  They used the direct DI calculation with DWF
and overlap which should have an inversion time comparable to the overlap fermion
we use. In our case, we used 176 configurations and 48 noise vectors to
calculate the high-mode part of the quark loop. In the case of
Engelhardt~\cite{Engelhardt:2012gd}, $\sim 468$ configurations were used each
with 1200 noise vectors, giving a relative error of 24\% at sink time $T = 10$,
comparable to our result.  However, this approach requires $\sim 66$ times as
many inversions as ours.  As for the JLQCD calculation~\cite{Oksuzian:2012rzb},
288 noise vectors were used on 50 configurations and the error is 2.54 times
larger than ours. Thus, to reach the same error as ours, it would take this
approach $\sim 11$ times as many inversions.  We attribute the efficiency
of our approach to the improvement of both the nucleon propagator and the quark
loop.

In the present work we have considered the statistical error only.
We will continue this work on the $32^3 \times 64$ dislocation-suppressing determinant ratio lattices with $a^{-1} =
1.37$ GeV and $m_{\pi} = 170$ and 250 MeV, as well as the finer $32^3 \times
64$ lattices with $a^{-1} = 2.31$ GeV and $m_{\pi} = 290$ MeV, in order to
extrapolate to the continuum limit and the physical sea pion mass in order to
address the systematic errors.

{\bf ACKNOWLEDGMENTS}

We thank RBC and UKQCD for sharing the DWF gauge configurations that we used in
the present work.  We thank Stefan Meinel for his fitting code.  
We also thank Andre Walker-Loud for showing us the mixed-action
partially quenched chiral perturbation expression in Eq.~(\ref{eq:chiral1}).
K.F. L. wishes to thank the Nuclear Theory Group at Lawrence Berkeley Lab. for their hospitality where part of the manuscript was written up.
This work is partially supported by DOE grants DE-FG05-84ER40154 and DE-FG02-00ER41132.
A. A. is partially supported by NSF CAREER grant PHY-1151648.  T. D. is
partially supported in part by MEXT Grant-in-Aid for Young Scientists (B)
(24740146). Z. L. is partially supported by NSFC under the Project 11105153.

%
%
\bibliographystyle{h-physrev3}
\bibliography{strangeness}

\end{document}